\documentclass[a4paper,11pt]{article}
\pdfoutput=1 
\usepackage{jcappub}

\usepackage{amsmath}
 
\usepackage{bm}
\usepackage{graphicx}
\usepackage{enumitem}
\usepackage{geometry}
\usepackage{xspace}
\usepackage{pdflscape}
\usepackage{placeins}
\usepackage{epstopdf}
\usepackage{comment}
\usepackage{xcolor}
\usepackage{soul}
\usepackage[makeroom]{cancel}

\makeatletter
\gdef\@fpheader{}
\g@addto@macro\bfseries{\boldmath}
\makeatother

\newcommand{\ie}{{i.e.~}}

\newcommand{\eg}{e.g.~}

\let\oldsqrt\sqrt
\def\sqrt{\mathpalette\DHLhksqrt}
\def\DHLhksqrt#1#2{%
\setbox0=\hbox{$#1\oldsqrt{#2\,}$}\dimen0=\ht0
\advance\dimen0-0.2\ht0
\setbox2=\hbox{\vrule height\ht0 depth -\dimen0}%
{\box0\lower0.4pt\box2}}

\newcommand{\order}[1]{\mathcal{O}\!\left(#1\right)}

\newcommand{\dd}{\mathrm{d}}
\newcommand{\ee}{e}

\newcommand{\sss}[1]{{\scriptscriptstyle{#1}}}
\newcommand{\boldmathsymbol}[1]{{\ensuremath{\boldsymbol{#1}}}}

\newcommand{\uPl}{\mathrm{Pl}}

\newcommand{\umax}{\mathrm{max}}

\newcommand{\ueff}{\mathrm{eff}}

\newcommand{\usssPl}{\sss{\uPl}}

\newcommand{\ud}{\mathrm{d}}

\newcommand{\calH}{\mathcal{H}}
\newcommand{\calP}{\mathcal{P}}

\newcommand{\GeV}{\mathrm{GeV}}

\newcommand{\Mpc}{\mathrm{Mpc}}

\newcommand{\cs}{c_{_\mathrm{S}}}

\newcommand{\Mp}{M_\usssPl}

\newcommand{\efolds}{$e$-folds}

\newcommand{\beq}{\begin{equation}}
\newcommand{\eeq}{\end{equation}}
\newcommand{\bea}{\begin{equation}\begin{aligned}}
\newcommand{\eea}{\end{aligned}\end{equation}}

\newlength{\wsingfig}
\newlength{\wdblefig}
\newlength{\wquadfig}
\newlength{\wtriplefig}
\newcommand{\Eq}[1]{Eq.~(\ref{#1})}
\newcommand{\Eqs}[1]{Eqs.~(\ref{#1})}
\newcommand{\Fig}[1]{Fig.~{\ref{#1}}}

\newcommand{\Refa}[1]{Ref.~{\cite{#1}}}
\newcommand{\Refs}[1]{Refs.~{\cite{#1}}}
\newcommand{\Sec}[1]{Sec.~\ref{#1}}

\newcommand{\App}[1]{Appendix~\ref{#1}}

\newcommand{\mPBH}{m_{\scriptscriptstyle{\mathrm{PBH}}}}
\newcommand{\rhoPBH}{\rho_{\scriptscriptstyle{\mathrm{PBH}}}}
\newcommand{\rhobarPBH}{\bar{\rho}_{\scriptscriptstyle{\mathrm{PBH}}}}
\newcommand{\OmegaPBH}{\Omega_{{\scriptscriptstyle{\mathrm{PBH}}}}}
\newcommand{\OmegaPBHf}{\Omega_{{\scriptscriptstyle{\mathrm{PBH}}},\mathrm{f}}}
\newcommand{\deltaPBH}{\delta_{\scriptscriptstyle{\mathrm{PBH}}}}
\newcommand{\zetaPBH}{\zeta_{\scriptscriptstyle{\mathrm{PBH}}}}

\newcommand{\rhoGW}{\rho_{\scriptscriptstyle{\mathrm{GW}}}}
\newcommand{\OmegaGW}{\Omega_{\scriptscriptstyle{\mathrm{GW}}}}
\newcommand{\OmegaGWtot}{\Omega_{\scriptscriptstyle{\mathrm{GW}},\mathrm{tot}}}

\newcommand{\Rcurv}{{\cal R}}

\subheader{}

\title{Gravitational waves from a universe filled with primordial black holes}

\author{Theodoros Papanikolaou,}
\author{Vincent Vennin,}
\author{David Langlois}

\affiliation{Laboratoire Astroparticule et Cosmologie, CNRS Universit\'e de Paris, 75013 Paris, France}

\emailAdd{theodoros.papanikolaou@apc.in2p3.fr}
\emailAdd{vincent.vennin@apc.univ-paris7.fr}
\emailAdd{langlois@apc.univ-paris7.fr}

\date{today}

\begin{document}

\sloppy

\abstract{
Ultra-light primordial black holes, with masses $\mPBH<10^9\mathrm{g}$, evaporate before big-bang nucleosynthesis and can therefore not be directly constrained. 
They can however be so abundant that they dominate the universe content for a transient period (before reheating the universe via Hawking evaporation). If this happens, they support large cosmological fluctuations at small scales, which in turn induce the production of gravitational waves through second-order effects. Contrary to the primordial black holes, those gravitational waves survive after evaporation, and can therefore be used to constrain such scenarios. In this work, we show that for induced gravitational waves not to lead to a backreaction problem, the relative abundance of black holes at formation, denoted $ \OmegaPBHf $, should be such that $ \OmegaPBHf <10^{-4}(\mPBH/10^9\mathrm{g})^{-1/4}$. In particular, scenarios where primordial black holes dominate right upon their formation time are all excluded (given that $\mPBH>10\, \mathrm{g}$ for inflation to proceed at  $\rho^{1/4}<10^{16}\mathrm{GeV}$). This sets the first constraints on ultra-light primordial black holes. 
}



\maketitle

\section{Introduction}
\label{sec:intro}
Primordial black holes (PBHs)~\cite{Carr:1974nx,1975ApJ...201....1C} are attracting increasing attention since they may play a number of important roles in Cosmology. They may indeed constitute part or all of the dark matter~\cite{Chapline:1975ojl}, they may explain the generation of large-scale structures through Poisson fluctuations~\cite{Meszaros:1975ef,Afshordi:2003zb}, they may provide seeds for supermassive black holes in galactic nuclei~\cite{1984MNRAS.206..315C, Bean:2002kx}, and they may also account for the progenitors of the black-hole merging events recently detected by the LIGO/VIRGO collaboration~\cite{LIGOScientific:2018mvr} through their gravitational wave emission, see \eg \Refs{Sasaki:2016jop, Abbott:2020mjq}. Other hints in favour of the existence of PBHs have been underlined, see for instance \Refa{2018PDU....22..137C} .

There are several constraints on the abundance of PBHs~\cite{Carr:2020gox}, ranging from micro-lensing constraints, dynamical constraints (such as constraints from the abundance of wide dwarfs in our local galaxy, or from the existence of a star cluster near the centres of ultra-faint dwarf galaxies), constraints from the cosmic microwave background due to the radiation released in PBH accretion, and constraints from the extragalactic gamma-ray background to which Hawking evaporation of PBHs contributes. However, all these constraints are restricted to certain mass ranges for the black holes, and no constraint applies to black holes with masses smaller than $\sim 10^{9}\mathrm{g}$, since those would Hawking evaporate before big-bang nucleosynthesis. 

Nonetheless, various scenarios have been proposed~\cite{Anantua:2008am,Zagorac:2019ekv,Martin:2019nuw,Inomata:2020lmk,Hooper:2019gtx} where ultra-light black holes are abundantly produced in the early universe, so abundantly that they might even dominate the energy budget of the universe for a transient period. By Hawking evaporating before big-bang nucleosynthesis takes place, those PBHs would leave no direct imprint (apart from possible Planckian relics~\cite{Markov:1984xd, Coleman:1991ku}). It thus seems rather frustrating that such a drastic change in the cosmological standard model, where an additional matter-dominated epoch driven by PBHs is introduced, and where reheating proceeds from PBH evaporation, cannot be constrained by the above-mentioned probes. This situation could however be improved by noting that a gas of gravitationally interacting PBHs is expected to emit gravitational waves, and that these gravitational waves would propagate in the universe until today, leaving an indirect imprint of the PBHs past existence.

The goal of this paper is therefore to compute the stochastic gravitational-wave background produced out of a gas of PBHs, if it constitutes the main component
of the universe. Since a gas of randomly distributed PBHs is associated with a density-fluctuation field, at second order in perturbation theory~\cite{Acquaviva:2002ud,Nakamura:2004rm}, these scalar fluctuations are expected to source the production of tensor perturbations~\cite{Ananda:2006af,Baumann:2007zm,Kohri:2018awv}, thus inducing a stochastic gravitational-wave background~\cite{Assadullahi:2009nf}. 

Let us note that there are several ways PBHs can be involved in the production of gravitational waves. First, the induction of gravitational waves can proceed from the primordial, large curvature perturbations that must have preceded (and given rise to) the existence of PBHs in the very early universe~\cite{Bugaev:2009zh, Saito:2008jc, Nakama:2015nea, Nakama:2016enz, Cai:2018dig, Yuan:2019udt, Kapadia:2020pir}. Second, the relic Hawking-radiated gravitons may also contribute to the stochastic gravitational-wave background~\cite{Anantua:2008am,Dong:2015yjs}. Third, gravitational waves are expected to be emitted by PBHs mergers~\cite{Nakamura:1997sm, Ioka:1998nz, Eroshenko:2016hmn, Raidal:2017mfl, Zagorac:2019ekv,Hooper:2020evu}. Here, we investigate a fourth effect, namely the production of gravitational waves induced by the large-scale density perturbations underlain by PBHs themselves. 
Contrary to the first effect mentioned above, more commonly studied, where PBHs and gravitational waves have a common origin (namely the existence of a large primordial curvature perturbation), in the problem at hand the gravitational waves are produced by the PBHs, via the gravitational potential they underlie. Let us also notice that since we make use of cosmological perturbation theory, we will restrict our analysis to scales larger than the mean separation distance between black holes, while the inclusion of smaller scales would require to resolve non-linear mechanisms such as merging, described in the third effect mentioned above.

As we will show, this fourth route is a very powerful one to constrain scenarios where the universe is transiently dominated by PBHs, since the mere requirement that the energy contained in the emitted gravitational waves does not overtake the one of the background (which would lead to an obvious backreaction problem), leads to tight constraints on the abundance of PBHs at the time they form. In particular, it excludes the possibility that PBHs dominate the universe upon their time of formation, independently of their mass. 

In practice, we consider that PBHs are initially randomly distributed in space, since recent works~\cite{Desjacques:2018wuu, Ali-Haimoud:2018dau, MoradinezhadDizgah:2019wjf} suggest that initial clustering is indeed negligible. We also assume that the mass distribution of PBHs is monochromatic, since it was shown to be the case in most formation mechanisms~\cite{Germani:2018jgr, MoradinezhadDizgah:2019wjf}. If PBHs form during the radiation era, their contribution to the total energy density
increases as an effect of the expansion. Therefore, if their initial abundance is sufficiently large, they dominate the universe content before they evaporate, and we compute the amount of gravitational waves produced during the PBH-dominated era. 

The paper is organised as follows. In \Sec{sec:PBH Distribution}, we explain how the initial PBH distribution can be modelled, and derive the gravitational potential it is associated with. In \Sec{sec:induced GWs basics}, we recall how the gravitational waves induced at second order from scalar perturbations can be computed, before applying these methods to the case of a PBH-dominated universe in \Sec{sec:Results}. There, we derive explicit constraints on the initial abundance of PBHs, as a function of their mass. We summarise our main results and conclude in \Sec{sec:Conclusions}. The paper finally contains several appendices where various technical aspects of the calculation are deferred.
\section{Gravitational potential of a gas of primordial black holes}
\label{sec:PBH Distribution}
In this section, we compute the power spectrum of the gravitational potential that is underlain by a gas of randomly distributed primordial black holes.
\subsection{Matter power spectrum}
\label{sec:matter power spectrum}
As explained in the introduction, we consider a gas of PBHs having all the same mass $\mPBH$, randomly distributed in space. This means that the probability distribution associated to the position of each black hole is uniform in space, and that the locations of several black holes are uncorrelated. In other words, their statistics is of the Poissonian type. This assumption neglects finite-size effects and the existence of an exclusion zone surrounding the position of each black hole, it is therefore not suited to describe length scales smaller than the Schwarzschild radius of the black holes. Moreover, in order to describe the PBH gas as a matter fluid sourcing a perturbatively small gravitational potential, our calculation has to be restricted to distances $r$ that are larger than the mean separation $\bar{r}$ between two neighbouring black holes. Below $\bar{r}$, the granularity of the PBH fluid becomes important. Since $\bar{r}$ is always larger than the Schwarzschild radius, it is sufficient to restrict the considerations below to $r>\bar{r}$. 

In \App{app:Poisson Statistics}, we show that the Poissonian approximation leads to the following real-space two-point function for the density contrast,
\bea
\label{eq:2pt:function:real:space:main:text}
\left\langle \frac{\delta  \rhoPBH (\bm{x})}{\rho_\mathrm{tot}}\frac{\delta  \rhoPBH (\bm{x}')}{\rho_\mathrm{tot}}\right\rangle = \frac{4}{3}\pi \left(\frac{\bar{r}}{a}\right)^3 \OmegaPBH^2\delta(\bm{x}-\bm{x}')\, ,
\eea
see \Eq{eq:2pt:function:real:space} (here, contrary to \Eq{eq:2pt:function:real:space}, $\bm{x}$ denotes comoving coordinates, hence the appearance of the scale factor $a$). In this expression, $\rhoPBH$ is the mass density of black holes,  $\rho_{\mathrm{tot}}$ is the overall mean energy density of the background, $\bar{r}$ can be expressed in terms of the mass and mean mass density of the black holes via $\bar{r}=\left(\frac{3\mPBH}{4\pi\rhobarPBH}\right)^{1/3}$, see \Eq{eq:app:rho_bar}, and $ \OmegaPBH \equiv \bar{\rho}_\mathrm{PBH}/\rho_{\mathrm{tot}}$ is the fractional energy density of the black holes.

Upon Fourier expanding the density contrast as
\bea
\label{eq:density:contrast:Fourier}
\frac{\delta  \rhoPBH (\bm{x})}{\rhobarPBH} = \int\frac{\dd ^3\bm{k}}{(2\pi)^{3/2}}\delta_{\bm{k}}(t)e^{i\boldmathsymbol{k}\cdot\boldmathsymbol{x}}\, ,
\eea
its power spectrum, $P_\delta(k)$, defined as $\langle \delta_{\bm{k}} \delta^{*}_{\bm{k}'} \rangle \equiv P_\delta(k) \delta(\bm{k}-\bm{k}')$,  
can be read off from plugging \Eq{eq:density:contrast:Fourier} into \Eq{eq:2pt:function:real:space:main:text}, and is given by
\bea
P_\delta(k)=\frac{4\pi}{3}\left(\frac{\bar{r}}{a}\right)^{3} \, .
\eea
The power spectrum is thus independent of $k$, a well-known result for Poissonian statistics. As explained above, the description of the PBH gas in terms of a continuous fluid is only valid at scales larger than the mean separation distance $\bar{r}$, which imposes an ultra-violet cutoff in the above power spectrum,
\bea
\label{k_UV}
k_{\mathrm{UV}} = \frac{a}{\bar{r}}\, .
\eea
In particular, it guarantees that the reduced power spectrum,
\bea
\label{eq:PowerSpectrum:PBH}
\calP_\delta(k) = \frac{k^3}{2\pi^2}P_\delta(k)= \frac{2}{3\pi} \left(\frac{k}{k_{\mathrm{UV}}}\right)^3\,,
\eea
is smaller than one since its maximal value is $\calP_\delta(k_{\mathrm{UV}}) = 2 /(3 \pi)\simeq 0.2$.
\subsection{Power spectrum of the gravitational potential}
\label{sec:CurvaturePowerSpectrum}
Our next task is to derive the power spectrum of the gravitational potential associated to PBHs at the onset of the PBH-dominated era. Since the Poissonian power spectrum for the density contrast derived in \Eq{eq:PowerSpectrum:PBH} holds at the time PBHs form, this implies to relate the initial PBH density contrast, computed in the radiation era, to the gravitational potential in the subsequent matter-dominated era. 

When PBHs are formed during the radiation era, their energy density is negligible with respect to the  energy density of the background, and the density contrast $\deltaPBH$ can thus be seen as an isocurvature perturbation~\cite{Inman:2019wvr}. This isocurvature perturbation then generates,  in the PBH-dominated era, a curvature perturbation, which we now compute. 

It is first convenient to introduce the uniform-energy-density curvature perturbation for the two components, namely~\cite{Wands:2000dp}
\bea
\zeta_\mathrm{r}=-\Phi+\frac14 \delta_\mathrm{r}
\eea
for the radiation fluid, where $\Phi$ is the Bardeen potential~\cite{Bardeen:1980kt}, and
\bea
\zetaPBH=-\Phi+\frac13 \deltaPBH
\eea
for the non-relativistic matter component, \ie the gas of PBHs. Let us see how these curvature perturbations evolve on super-Hubble ($k\ll \mathcal{H}$, where $\mathcal{H}$ is the comoving Hubble parameter) and sub-Hubble ($k\gg \mathcal{H}$) scales.

On super-Hubble scales, $\zeta_\mathrm{r}$ and $\zetaPBH$ are separately conserved~\cite{Wands:2000dp},  as is the isocurvature perturbation defined by
\bea
S=3\left(\zetaPBH-\zeta_\mathrm{r}\right)\,.
\eea
By contrast, the total curvature perturbation, 
\beq
\zeta=-\Phi+\frac {\delta_{\mathrm{tot}}}{3(1+w)}=\frac{4}{4+3s}\zeta_\mathrm{r}+\frac{3s}{4+3s}\zetaPBH\,, \quad {\rm with} \quad s\equiv \frac{a}{a_\mathrm{d}}\,,
\eeq
evolves from its initial value $\zeta_\mathrm{r}$, deep in the radiation era, to $\zetaPBH$, deep in the PBH era. In this expression, $w$ is the equation-of-state parameter, and $a_\mathrm{d}$ denotes the value of the scale factor $a$ at the time PBHs start dominating. As a consequence, in the PBH-dominated era, $\zeta\simeq \zetaPBH = \zeta_{\mathrm{r}}+S/3$. Since $S$ is conserved, it can be evaluated at formation time $t_\mathrm{f}$. Furthermore, the isocurvature perturbation can be identified with $\deltaPBH(t_\mathrm{f})$, which we have computed in the previous section, assuming implicitly a uniform radiation energy density in the background. Indeed, in the following, we concentrate on the PBH contribution and ignore the usual adiabatic contribution (associated to the radiation fluid), which is negligible on the scales we are interested in, hence one simply has
\bea
\label{eq:zeta:delta:superH}
\zeta\simeq \frac13 \deltaPBH(t_\mathrm{f})\quad \mathrm{if}\quad k\ll {\calH}\,.
\eea
One can then use the property that $\zeta\simeq -\Rcurv$ on super-Hubble scales (see e.g. \Refa{Wands:2000dp}), where $\Rcurv$ is the comoving curvature perturbation defined by
\bea
\label{eq:zeta:Bardeen}
\Rcurv  = \frac{2}{3}\frac{{\Phi}^\prime/\calH+\Phi}{1+w}+\Phi\, .
\eea
During a matter-dominated era, such as the one driven by PBHs, ${\Phi}^\prime $ can be neglected since it is proportional to the decaying mode, so we get  $\Rcurv=-\zeta=(5/3)\Phi$. Combining with \Eq{eq:zeta:delta:superH}, this  implies that
\bea
\label{eq:Phi:delta:superH}
\Phi\simeq -\frac15 \deltaPBH(t_\mathrm{f})\quad \mathrm{if}\quad k\ll {\calH}\,.
\eea
On sub-Hubble scales, one can determine the evolution of $\deltaPBH$ by solving its equation of motion~\cite{Meszaros:1974tb}, 
\bea
\frac{\dd^2 \deltaPBH}{\dd s^2}+\frac{2+3s}{2s(s+1)}\frac{\dd \deltaPBH}{\dd s}-\frac{3}{2s (s+1)}\deltaPBH=0\,,
\eea
the dominant solution of which is given by
\bea
\deltaPBH = \frac{2+3s}{2+3s_\mathrm{f}} \deltaPBH(t_\mathrm{f})\, .
\eea
Let us stress that this formula is valid at all scales, and that, since it does not involve the wavenumber $k$, it implies that the statistical distribution of PBHs remains Poissonian, \ie $\calP_{\deltaPBH}\propto k^3$ even after formation time~\cite{Desjacques:2018wuu, Ali-Haimoud:2018dau, MoradinezhadDizgah:2019wjf}. Deep in the PBH-dominated era, neglecting $s_\mathrm{f}$, it gives rise to $\deltaPBH\simeq 3s\,  \deltaPBH(t_\mathrm{f})/2$. On sub-Hubble scales, the relation between the Bardeen potential and the density contrast does not depend on the slicing in which the density contrast is defined, and in a matter-dominated era, it takes the form
\bea
\label{eq:delta:Phi:MD}
\deltaPBH =  -\frac{2}{3} \left(\frac{k}{\calH}\right)^2\Phi \, .
\eea
Plugging the solution we have obtained for $\deltaPBH$ into this formula, one obtains
\bea
\label{eq:Phi:delta:subH}
\Phi\simeq -\frac{9}{4}\left(\frac{\calH_{\mathrm{d}}}{k}\right)^2\, \deltaPBH(t_\mathrm{f})
\quad \mathrm{if}\quad k\gg {\calH}_{\mathrm{d}}\, .
\eea
From \Eq{eq:Phi:delta:superH} and \Eq{eq:Phi:delta:subH}, one can see that, both on sub- and super-Hubble scales, the Bardeen potential is constant during the PBH era, in agreement with the expected behaviour in a matter-dominated epoch. Using a crude interpolation between the two expressions, one obtains
\beq
\label{eq:PhiPBH:deltaPBH}
\Phi\simeq -\left(5+\frac{4}{9}\frac{k^2}{\calH_{\mathrm{d}}^2}\right)^{-1}\, \delta_{\rm PBH}(t_\mathrm{f})\,.
\eeq
Combining \Eqs{eq:PowerSpectrum:PBH} and~\eqref{eq:PhiPBH:deltaPBH}, the power spectrum for the Bardeen potential is finally given by
\bea
\label{eq:PowerSpectrum:Phi:PBHdom}
\calP_\Phi (k) = \frac{2}{3\pi}\left(\frac{k}{k_\mathrm{UV}}\right)^3 \left(5+\frac{4}{9}\frac{k^2}{\calH_{\mathrm{d}}^2}\right)^{-2}\, ,
\eea
where we have made use of \Eq{k_UV} to replace $a/\bar{r}$ by $k_{\mathrm{UV}}$. Notice that, since $\bar{r}\propto a$, $k_{\mathrm{UV}}$ is a fixed comoving scale. From \Eq{eq:PowerSpectrum:Phi:PBHdom}, one can see that $\calP_\Phi$ is made of two branches: when $k\ll \calH_{\mathrm{d}}$, $\calP_\Phi \propto k^3$, while $\calP_\Phi \propto 1/k$ when $k\gg \mathcal{H}_\mathrm{d}$. It reaches a maximum when $k \sim \calH_\mathrm{d}$, where $\calP_\Phi$ is of order $(\calH_\mathrm{d}/k_{\mathrm{UV}})^3$. 
\section{Scalar-induced gravitational waves}
\label{sec:induced GWs basics}
Having determined the gravitational potential associated with the gas of PBHs,
let us now work out the gravitational waves that this gravitational potential induces.
\subsection{Gravitational waves at second order}
\label{sec:2nd:order:GWs}
Although tensor modes are gauge invariant at first order in perturbation theory, this does not hold at second order~\cite{Hwang:2017oxa, Tomikawa:2019tvi}. This means that, a priori, one needs to specify in which slicing the gravitational waves are observed, \ie which coordinate system is employed to perform the detection. This depends on the specifics of the detection apparatus. Recently, it has been shown that the gauge dependence of the result disappears if gravitational waves are emitted during a radiation era~\cite{DeLuca:2019ufz, Yuan:2019fwv, Inomata:2019yww}. Although we study the case where gravitational waves are emitted during a PBH-dominated era, hence a matter era, for which the question is more subtle, we are not aiming at deriving observable predictions, but rather at investigating a backreaction problem, which we assume bears little dependence on the gauge: if the energy density carried by gravitational waves becomes comparable with the one of the background, one expects perturbation theory to break down in any gauge. 

In practice, we choose to follow \Refs{Ananda:2006af,Baumann:2007zm,Kohri:2018awv,Espinosa:2018eve} and to work in the Newtonian gauge. Adding to the linearly-perturbed Friedmann-Lema\^itre-Robertson-Walker metric in the Newtonian gauge the second-order tensor perturbation $h_{ij}$ (with a factor $1/2$ as is standard in the literature)\footnote{The first-order tensor perturbation is ignored here as we concentrate on gravitational waves generated by scalar perturbations at second order, but can be added to the contribution computed in our work, for instance to include the gravitational waves produced during inflation via the usual mechanism.}, we obtain the total metric
\bea
\label{metric decomposition with tensor perturbations}
\mathrm{d}s^2 = a^2(\eta)\left\lbrace-(1+2\Phi)\mathrm{d}\eta^2  + \left[(1-2\Phi)\delta_{ij} + \frac{h_{ij}}{2}\right]\mathrm{d}x^i\mathrm{d}x^j\right\rbrace \, .
\eea
The tensor perturbation can be Fourier expanded according to
\beq
\label{h_ij Fourier decomposition}
h_{ij}(\eta,\boldmathsymbol{x}) = \int \frac{\mathrm{d}^3\boldmathsymbol{k}}{\left(2\pi\right)^{3/2}} \left[h^{(+)}_\boldmathsymbol{k}(\eta)e^{(+)}_{ij}(\boldmathsymbol{k}) + h^{(\times)}_\boldmathsymbol{k}(\eta)e^{(\times)}_{ij}(\boldmathsymbol{k}) \right]e^{i\boldmathsymbol{k}\cdot\boldmathsymbol{x}},
\eeq
with the polarisation tensors $e^{(+)}_{ij}$ and $e^{(-)}_{ij}$  
defined as
\begin{eqnarray}
e^{(+)}_{ij}(\boldmathsymbol{k}) = \frac{1}{\sqrt{2}}\left[e_i(\boldmathsymbol{k})e_j(\boldmathsymbol{k}) - \bar{e}_i(\boldmathsymbol{k})\bar{e}_j(\boldmathsymbol{k})\right], \\ 
e^{(\times)}_{ij}(\boldmathsymbol{k}) = \frac{1}{\sqrt{2}}\left[e_i(\boldmathsymbol{k})\bar{e}_j(\boldmathsymbol{k}) + \bar{e}_i(\boldmathsymbol{k})e_j(\boldmathsymbol{k})\right],
\end{eqnarray}
where $e_i(\boldmathsymbol{k})$ and $\bar{e}_i(\boldmathsymbol{k})$ are two three-dimensional vectors, such that $\lbrace e_i(\boldmathsymbol{k}), \bar{e}_i(\boldmathsymbol{k}), \boldmathsymbol{k}/k \rbrace$ forms an orthonormal basis. This implies that the polarisation tensors satisfy $e^{(+)}_{ij}e^{(+)}_{ij} = e^{(\times)}_{ij}e^{(\times)}_{ij} =1 , e^{(+)}_{ij}e^{(\times)}_{ij}=0$. The equation of motion for the tensor modes is given by~ \cite{Ananda:2006af,Baumann:2007zm,Kohri:2018awv}
\beq
\label{Tensor Eq. of Motion}
h_\boldmathsymbol{k}^{s,\prime\prime} + 2\mathcal{H}h_\boldmathsymbol{k}^{s,\prime} + k^{2}h^s_\boldmathsymbol{k} = 4 S^s_\boldmathsymbol{k}\, ,
\eeq
where 
$s = (+), (\times)$ and the source function $S^s_\boldmathsymbol{k}$ is given by
\beq
\label{eq:Source:def}
S^s_\boldmathsymbol{k}  = \int\frac{\mathrm{d}^3 \boldmathsymbol{q}}{(2\pi)^{3/2}}e^s_{ij}(\boldmathsymbol{k})q_iq_j\left[2\Phi_\boldmathsymbol{q}\Phi_\boldmathsymbol{k-q} + \frac{4}{3(1+w)}(\mathcal{H}^{-1}\Phi_\boldmathsymbol{q} ^{\prime}+\Phi_\boldmathsymbol{q})(\mathcal{H}^{-1}\Phi_\boldmathsymbol{k-q} ^{\prime}+\Phi_\boldmathsymbol{k-q}) \right]\, .
\eeq
The source being quadratic in $\Phi$, it is a second-order quantity, and so are the tensor modes. In \Eq{eq:Source:def}, the contraction  $e^s_{ij}(\boldmathsymbol{k})q_iq_j \equiv e^s(\boldmathsymbol{k},\boldmathsymbol{q})$ can be expressed in terms of the spherical coordinates $(q,\theta,\varphi)$ of the vector $\bm{q}$ in the basis $\lbrace e_i(\boldmathsymbol{k}), \bar{e}_i(\boldmathsymbol{k}), \boldmathsymbol{k}/k \rbrace$, 
\beq
e^s(\boldmathsymbol{k},\boldmathsymbol{q})=
\begin{cases}
\frac{1}{\sqrt{2}}q^2\sin^2\theta\cos 2\varphi \mathrm{\;for\;} s= (+)\\
\frac{1}{\sqrt{2}}q^2\sin^2\theta\sin 2\varphi  \mathrm{\;for\;} s= (\times)
\end{cases}
\, .
\eeq
In the absence of anisotropic stress, if the speed of sound is given by $\cs^2=w$, the equation of motion for the Bardeen potential reads~\cite{1992PhR...215..203M}
\bea
\label{Bardeen potential 2}
\Phi_\boldmathsymbol{k}^{\prime\prime} + \frac{6(1+w)}{1+3w}\frac{1}{\eta}\Phi_\boldmathsymbol{k}^{\prime} + wk^2\Phi_\boldmathsymbol{k} =0\, .
\eea
Introducing $x\equiv k \eta$ and $\lambda\equiv (5+3w)/(2+6w)$, this can be solved in terms of the Bessel functions $J_\lambda$ and $Y_\lambda$,
\beq\label{Bardeen Potential - Exact Solution}
\Phi_\boldmathsymbol{k}(\eta) = \frac{1}{x^{\lambda}}\left[C_1(k)J_\mathrm{\lambda}\left(\sqrt{w}x\right) + C_2(k)Y_\mathrm{\lambda}\left(\sqrt{w}x\right)\right],
\eeq
where $C_1(k)$ and $C_2(k)$ are two integration constants. On super sound-horizon scales, \ie when $\sqrt{w}\vert x\vert \ll 1$, this solution features a constant mode and a decaying mode (when $w=0$, this is valid at all scales). By considering the Bardeen potential after it has spent several \efolds~above the sound horizon, the decaying mode can be neglected, and one can write $\Phi_\boldmathsymbol{k}(\eta) = T_\Phi(x) \phi_\boldmathsymbol{k}$, where $\phi_\boldmathsymbol{k}$ is the value of the Bardeen potential at some reference initial time (which here we take to be the time at which PBHs start dominating, $x_\ud$) and $T_\Phi(x)$ is a transfer function, defined as the ratio of the dominant mode between the times $x$ and $x_\ud$. This allows one to rewrite \Eq{eq:Source:def} as
\beq
\label{Source}
S^s_\boldmathsymbol{k}  =
\int\frac{\mathrm{d}^3 q}{(2\pi)^{3/2}}e^{s}(\boldmathsymbol{k},\boldmathsymbol{q})F(\boldmathsymbol{q},\boldmathsymbol{k-q},\eta)\phi_\boldmathsymbol{q}\phi_\boldmathsymbol{k-q}\, ,
\eeq
where one has introduced
\bea
\label{F}
F(\boldmathsymbol{q},\boldmathsymbol{k-q},\eta) & = 2T_\Phi(q\eta)T_\Phi\left(|\boldmathsymbol{k}-\boldmathsymbol{q}|\eta\right) 
\\  & \kern-2em + \frac{4}{3(1+w)}\left[\mathcal{H}^{-1}qT_\Phi^{\prime}(q\eta)+T_\Phi(q\eta)\right]\left[\mathcal{H}^{-1}\vert\boldmathsymbol{k}-\boldmathsymbol{q}\vert T_\Phi^{\prime}\left(|\boldmathsymbol{k}-\boldmathsymbol{q}|\eta\right)+T_\Phi\left(|\boldmathsymbol{k}-\boldmathsymbol{q}|\eta\right)\right],
\eea
which only involves the transfer function $T_\Phi$.

A formal solution to \Eq{Tensor Eq. of Motion} is obtained with the Green's function formalism,
\bea
\label{tensor mode function}
a(\eta)h^s_\boldmathsymbol{k} (\eta)  =4 \int^{\eta}_{\eta_\mathrm{d}}\mathrm{d}\bar{\eta}\,  g_\boldmathsymbol{k}(\eta,\bar{\eta})a(\bar{\eta})S^s_\boldmathsymbol{k}(\bar{\eta}),
\eea
where the Green's function $g_\boldmathsymbol{k}(\eta,\bar{\eta})$ is given by $g_\boldmathsymbol{k}(\eta,\bar{\eta})=G_\boldmathsymbol{k}(\eta,\bar{\eta})\Theta(\eta-\bar{\eta})$. In this expression, $\Theta$ is the Heaviside step function, and $G_{\bm{k}}(\eta,\bar{\eta})$ is the solution of the homogeneous equation 
\beq
\label{Green function equation}
G_\boldmathsymbol{k}^{\prime\prime}  + \left( k^{2} - \frac{a^{\prime\prime}}{a}\right)G_\boldmathsymbol{k} = 0\, ,
\eeq
where a prime denotes derivation with respect to the first argument $\eta$, and with initial conditions $ \lim_{\eta\to \bar{\eta}}G_\boldmathsymbol{k}(\eta,\bar{\eta}) = 0$ and $ \lim_{\eta\to \bar{\eta}}G^\prime_\boldmathsymbol{k}(\eta,\bar{\eta})=1$.  The above equation can be solved analytically in terms of Bessel functions and the solution is:
\beq\label{Green function analytical}
kG_\boldmathsymbol{k}(\eta,\bar{\eta}) =\frac{\pi}{2} \sqrt{x\bar{x}}\left[Y_\mathrm{\nu}(x)J_\mathrm{\nu}(\bar{x})  - J_\mathrm{\nu}(x)Y_\mathrm{\nu}(\bar{x})\right],
\eeq
where $\nu = \frac{3(1-w)}{2(1+3w)}$. Since $G_\boldmathsymbol{k}(\eta,\bar{\eta})$ depends only on $k$, from now on it will be noted as $G_k(\eta,\bar{\eta})$.
\subsection{The stress-energy tensor of gravitational waves}
\label{sec:Stress:Energy:GW}
Now that we have derived the amplitude of the gravitational waves induced by scalar perturbations, let us study the energy density they give rise to. Following closely \Refa{Maggiore:1999vm}, we consider only the contribution from small-scale perturbations, \ie scales $\lambda$ that are much smaller than the scale characterising the background metric $L_\mathrm{B}$. By coarse graining perturbations below the intermediate scale $\ell$ such that $\lambda\ll \ell \ll L_\mathrm{B}$, the effective stress-energy tensor of gravitational waves reads~\cite{Maggiore:1999vm}
\beq
\label{t_munu 1}
t_\mathrm{\mu\nu}=-\Mp^2\, \overline{ \left(R^{(2)}_\mathrm{\mu\nu} - \frac{1}{2}\bar{g}_\mathrm{\mu\nu}R^{(2)}\right)},
\eeq
where 
 $\bar{g}_\mathrm{\mu\nu}$ is the background metric, 
$R^{(2)}_\mathrm{\mu\nu}$ is the second-order Ricci tensor and  $R^{(2)} = \bar{g}^\mathrm{\mu\nu}R^{(2)}_\mathrm{\mu\nu}$ its trace. 
The overall bar refers to the coarse-graining procedure.

The physical modes contained in $t_\mathrm{\mu\nu}$ can be extracted either by specifying a gauge, as for instance the transverse-traceless gauge where $\partial_\mathrm{\beta}h^\mathrm{\alpha\beta}=0$ and  $h=\bar{g}^\mathrm{\alpha\beta}h_\mathrm{\alpha\beta}=0$,
 or in a gauge-invariant way by using space-time averages~\cite{Maggiore:1999vm} (see also Appendix of \Refa{Isaacson:1968zza}). Both approaches coincide on sub-Hubble scales where space time is effectively flat, and where the 0-0 component of $t_{\mu\nu}$ reads
\bea
\label{rho_GW effective}
 \rhoGW (\eta,\boldmathsymbol{x}) = t_{00} &=  \frac{\Mp^2}{32 a^2}\, \overline{\left(\partial_\eta h_\mathrm{\alpha\beta}\partial_\eta h^\mathrm{\alpha\beta} +  \partial_{i} h_\mathrm{\alpha\beta}\partial^{i}h^\mathrm{\alpha\beta} \right)}\, ,
\eea
which is simply the sum of a kinetic term and a gradient term. 

In the case of a free wave [\ie in the absence of a source term in \Eq{Tensor Eq. of Motion}], these two contributions are identical, since the energy is equipartitioned between its kinetic and gradient components. In the present case however, in \App{rho_GW}, we show that the source term ``forces'' the amplitude of gravitational waves towards a constant solution, which highly suppresses the kinetic contribution compared to the gradient contribution. In this regime, only the gradient energy remains and \Eq{rho_GW effective} leads to 
\beq\label{rho_GW_effective MD}
\begin{split}
\left\langle \rhoGW (\eta,\boldmathsymbol{x}) \right\rangle & = t_{00}  \simeq \sum_{s=+,\times}\frac{\Mp^2}{32a^2}\overline{\left\langle\left(\nabla h^{s}_\mathrm{\alpha\beta}\right)^2\right \rangle }
   \\ & =   \frac{\Mp^2}{32a^2 \left(2\pi\right)^3} \sum_{s=+,\times} \int\mathrm{d}^3\boldmathsymbol{k}_1 \int\mathrm{d}^3\boldmathsymbol{k}_2\,  k_1 k_2 \overline{  \left\langle h^{s}_{\boldmathsymbol{k}_1}(\eta)h^{s,*}_{\boldmathsymbol{k}_2}(\eta)\right\rangle} e^{i(\boldmathsymbol{k}_1-\boldmathsymbol{k}_2)\cdot \boldmathsymbol{x}}\, .
 \end{split}
\eeq
In this expression, the bar denotes averaging over the  sub-horizon oscillations of the tensor field, which is done in order to only extract the envelope of the gravitational-wave spectrum at those scales and brackets mean an ensemble average.
\subsection{The tensor power spectrum at second order}
\label{sec:tensor:power:spectrum}
From the above expression, it is clear that our next step is to derive the two-point correlation function of the tensor field, $\langle h^r_{\boldmathsymbol{k}_1}(\eta)h^{s,*}_{\boldmathsymbol{k}_2}(\eta)\rangle$. As we will now show, it is of the form
\bea
\label{tesnor power spectrum definition}
\langle h^r_{\boldmathsymbol{k}_1}(\eta)h^{s,*}_{\boldmathsymbol{k}_2}(\eta)\rangle \equiv \delta^{(3)}(\boldmathsymbol{k}_1 - \boldmathsymbol{k}_2) \delta^{rs} \frac{2\pi^2}{k^3_1}\mathcal{P}_{h}(\eta,k_1),
\eea
where $\mathcal{P}_{h}(\eta,k)$ is the tensor power spectrum. According to \Eq{tensor mode function}, the two-point function of the tensor fluctuation can indeed be expressed in terms of the two-point function of the source,
\beq\label{h^rh^s analytic 1}
\langle h^r_{\boldmathsymbol{k}_1}(\eta)h^{s,*}_{\boldmathsymbol{k}_2}(\eta)\rangle = \frac{16}{a^2(\eta)}\int_{\eta_\mathrm{d}}^{\eta}\mathrm{d}\bar{\eta}_1 G_{k_1}(\eta,\bar{\eta}_1)a(\bar{\eta}_1)\int_{\eta_\mathrm{d}}^{\eta}\mathrm{d}\bar{\eta}_2G_{k_2}(\eta,\bar{\eta}_2)a(\bar{\eta}_2)\langle S^r_{\boldmathsymbol{k}_1}(\bar{\eta}_1)S^{s,*}_{\boldmathsymbol{k}_2}(\bar{\eta}_2)\rangle,
\eeq
where the source correlator can be derived from \Eq{Source}, leading to 
\beq\label{Source correlator 1}
\begin{split}
\langle S^r_{\boldmathsymbol{k}_1}(\bar{\eta}_1)S^{s,*}_{\boldmathsymbol{k}_2}(\bar{\eta}_2)\rangle & = \int\frac{\mathrm{d}^3 q_1}{(2\pi)^{3/2}}e^r(\boldmathsymbol{k}_1,\boldmathsymbol{q}_1)F(\boldmathsymbol{q}_1,\boldmathsymbol{k}_1-\boldmathsymbol{q}_1,\bar{\eta}_1) \\ & \kern -3em
\times \int\frac{\mathrm{d}^3 q_2}{(2\pi)^{3/2}}e^s(\boldmathsymbol{k}_2,\boldmathsymbol{q}_2)F^{*}(\boldmathsymbol{q}_2,\boldmathsymbol{k}_2-\boldmathsymbol{q}_2,\bar{\eta}_2)\langle \phi_{\boldmathsymbol{q}_1}\phi_{\boldmathsymbol{k}_1-\boldmathsymbol{q}_1} \phi^{*}_{\boldmathsymbol{q}_2}\phi^{*}_{\boldmathsymbol{k}_2-\boldmathsymbol{q}_2} \rangle .
\end{split}
\eeq
Making use of Wick theorem, the four-point correlator $\langle \phi_{\boldmathsymbol{q}_1}\phi_{\boldmathsymbol{k}_1-\boldmathsymbol{q}_1} \phi_{\boldmathsymbol{q}_2}\phi_{\boldmathsymbol{k}_2-\boldmathsymbol{q}_2} \rangle$ has two non-vanishing contractions for $k_1\neq 0$ and $k_2\neq 0$, namely 
\beq
\label{eq:phi:correlator}
\langle \phi_{\boldmathsymbol{q}_1}\phi_{\boldmathsymbol{k}_1-\boldmathsymbol{q}_1} \phi^{*}_{\boldmathsymbol{q}_2}\phi^{*}_{\boldmathsymbol{k}_2-\boldmathsymbol{q}_2} \rangle = \langle \phi_{\boldmathsymbol{q}_1}\phi^{*}_{\boldmathsymbol{k}_2-\boldmathsymbol{q}_2}\rangle\langle\phi_{\boldmathsymbol{k}_1-\boldmathsymbol{q}_1}\phi^{*}_{\boldmathsymbol{q}_2} \rangle + \langle \phi_{\boldmathsymbol{q}_1}\phi^{*}_{\boldmathsymbol{q}_2}\rangle\langle\phi_{\boldmathsymbol{k}_1-\boldmathsymbol{q}_1}\phi^{*}_{\boldmathsymbol{k}_2-\boldmathsymbol{q}_2} \rangle.
\eeq
These two terms yield the same contribution in \Eq{Source correlator 1}, which can be seen by performing the change of integration variable $\boldmathsymbol{q}_2\rightarrow \boldmathsymbol{k}_2-\boldmathsymbol{q}_2$.\footnote{The fact that $\langle \phi_{\boldmathsymbol{q}_1}\phi_{\boldmathsymbol{k}_1-\boldmathsymbol{q}_1} \phi^{*}_{\boldmathsymbol{q}_2}\phi^{*}_{\boldmathsymbol{k}_2-\boldmathsymbol{q}_2} \rangle$ remains unchanged when exchanging $\boldmathsymbol{q}_2$ and $\boldmathsymbol{k}_2-\boldmathsymbol{q}_2$ is obvious from \Eq{eq:phi:correlator}. In the same way, the fact that $F$ is symmetrical upon its two first arguments can be clearly seen in \Eq{F}. Finally, since $e^s(\boldmathsymbol{k},\boldmathsymbol{q})$ only involves scalar products of $\boldmathsymbol{q}$ with vectors orthogonal to $\boldmathsymbol{k}$, it is also clear that $e^s(\boldmathsymbol{k}_2,\boldmathsymbol{q}_2)=e^s(\boldmathsymbol{k}_2,\boldmathsymbol{k}_2-\boldmathsymbol{q}_2)$.}
One can therefore compute one such contribution only, and simply multiply the result by $2$.

In the PBH-dominated era,  the two-point correlation function of the Bardeen potential is related to the power spectrum~\eqref{eq:PowerSpectrum:Phi:PBHdom}  via
\bea
\label{curvature power spectrum vs phi_k}
\langle\phi_{\boldmathsymbol{k}_1}\phi^{*}_{\boldmathsymbol{k}_2}\rangle = \delta(\boldmathsymbol{k}_1-\boldmathsymbol{k}_2)\frac{2\pi^2}{k^3_1}\mathcal{P}_\Phi(k_1) .
\eea
Combining the above results, \Eq{Source correlator 1} gives rise to
\bea
\label{Source correlator 2}
\langle S^r_{\boldmathsymbol{k}_1}(\bar{\eta}_1)S^{s,*}_{\boldmathsymbol{k}_2}(\bar{\eta}_2)\rangle & = \pi \delta^{(3)}(\boldmathsymbol{k}_1 -\boldmathsymbol{k}_2)  
 \int \mathrm{d}^3 \boldmathsymbol{q}_1e^r(\boldmathsymbol{k}_1,\boldmathsymbol{q}_1)e^s(\boldmathsymbol{k}_1,\boldmathsymbol{q}_1) 
  \\ &
  F(\boldmathsymbol{q}_1,\boldmathsymbol{k}_1-\boldmathsymbol{q}_1,\bar{\eta}_1) F^{*}(\boldmathsymbol{q}_1,\boldmathsymbol{k}_1-\boldmathsymbol{q}_1,\bar{\eta}_2) \frac{\mathcal{P}_\Phi(q_1)}{q^3_1}\frac{\mathcal{P}_\Phi(|\boldmathsymbol{k}_1-\boldmathsymbol{q}_1|)}{|\boldmathsymbol{k}_1-\boldmathsymbol{q}_1|^3}.
\eea
It is then convenient to re-write the above integral in terms of the two auxiliary variables $u = |\boldmathsymbol{k}_1 - \boldmathsymbol{q}_1|/k_1$ and $v = q_1/k_1$. In the orthonormal basis $\lbrace e_i(\boldmathsymbol{k}_1), \bar{e}_i(\boldmathsymbol{k}_1), \boldmathsymbol{k}_1/k_1 \rbrace$, let $(q_1,\theta,\phi)$ be the spherical coordinates of the vector $\boldmathsymbol{q}_1$. Applying the law of cosines (also known as Al Kashi's theorem) to  the triangle formed of the vectors $\boldmathsymbol{k} _1$, $\boldmathsymbol{q} _1$ and $\boldmathsymbol{k} _1 - \boldmathsymbol{q} _1$, one finds $\cos\theta=(1+v^2-u^2)/2v$, while one simply has $q_1=k_1 v$. The integral over $\boldmathsymbol{q}_1$ can thus be written as
\beq
\int_{\mathbb{R}^3}\mathrm{d}^3\boldmathsymbol q_1 =  k^3_1\int_{0}^{\infty}\mathrm{d}v\, v^2  \int_{|1-v|}^{1+v}\mathrm{d}u\,   \frac{u}{v}   \int_0^{2\pi}\mathrm{d}\phi.
\eeq
Then, noticing that $F(\boldmathsymbol{q},\boldmathsymbol{k-q},\eta)$ depends only on the modulus of its first two arguments, see \Eq{F}, and given that, by construction, $\vert \boldmathsymbol{q}_1-\boldmathsymbol{k}_1\vert = k_1 u$ does not depend on $\phi$, the integral over $\phi$ in \Eq{Source correlator 2} can be performed independently, and one finds
\beq
\int_0^{2\pi}\mathrm{d}\phi \, e^r(\boldmathsymbol{k}_1,\boldmathsymbol{q}_1)e^s(\boldmathsymbol{k}_1,\boldmathsymbol{q}_1) = \frac{k^4_1}{2}v^4\left[1-\frac{(1+v^2-u^2)^2}{4v^2}\right]^2\pi\,  \delta^{rs}.
\eeq
Combining the above results, the two-point function of the tensor field can be cast in the form of \Eq{tesnor power spectrum definition}, where
 the tensor power spectrum is given by
\bea
\label{Tensor Power Spectrum}
\mathcal{P}_h(\eta,k) = 4\int_{0}^{\infty} \mathrm{d}v\int_{|1-v|}^{1+v}\mathrm{d}u \left[ \frac{4v^2 - (1+v^2-u^2)^2}{4uv}\right]^{2}I^2(u,v,x)\mathcal{P}_\Phi(kv)\mathcal{P}_\Phi(ku)\,,
\eea
with
\bea
\label{I function}
I(u,v,x) = \int_{x_\mathrm{d}}^{x} \mathrm{d}\bar{x}\, \frac{a(\bar{x})}{a(x)}\, k\, G_{k}(x,\bar{x}) F_k(u,v,\bar{x}).
\eea
In this expression, $x=k\eta$ and we use the notation  $F_{k}(u,v,\eta)\equiv  F(k ,|\boldmathsymbol{k}-\boldmathsymbol{q}|,\eta)$.

From \Eq{F}, the function $F_k(u,v,\bar{x})$ in a matter-dominated era reads
\bea
\label{f function - w =0}
F_k(u,v,\bar{x}) & =  \frac{10}{3}\, T_\Phi(u\bar{x})\, T_\Phi(v\bar{x})\,,
\eea
where we have used the property that $T_\Phi$ is constant in the PBH-dominated era.
Let us note that, in \Eq{Tensor Power Spectrum}, the tensor power spectrum is given as a convolution product of the gravitational-potential power spectrum at the scales $\boldmathsymbol{q}_1$ and $\boldmathsymbol{q}_2$ such that $\boldmathsymbol{q}_1+\boldmathsymbol{q}_2=\boldmathsymbol{k}$. According to the discussion in \Sec{sec:matter power spectrum}, scalar fluctuations above the UV cutoff should be discarded, which can be done by setting $\calP_\Phi(q)=0$ for $q>k_\mathrm{UV}$. Since $k<q_1+q_2$, this implies that  $\calP_h(k)=0$ for  $k>2 k_\mathrm{UV}$, so up to a factor $2$, the UV cutoff also applies to tensor modes.

Following a similar calculation as the one for the power spectrum, the energy density contained in gravitational waves, and given by \Eq{rho_GW_effective MD}, can also be derived. Defining $\OmegaGW(\eta,k)$ through the relation
\bea
\label{eq:OmegaGW:def}
\left\langle  \rhoGW (\eta,\boldmathsymbol{x}) \right\rangle \equiv \rho_\mathrm{tot} \int \OmegaGW(\eta,k)  \dd\ln k\, ,
\eea
one obtains (see \App{rho_GW} for further details)
\bea
\label{Omega_GW}
 \OmegaGW (\eta,k) =  \frac{1}{48}\left[\frac{k}{\calH(\eta)}\right]^{2}\overline{\mathcal{P}}_{h}(\eta,k) .
\eea
\section{Constraints on the abundance of primordial black holes}
\label{sec:Results}
In this section, we carry out the calculational program derived in \Sec{sec:induced GWs basics} and compute the energy density contained in the induced gravitational waves. The two parameters of the problem are the mass of the PBHs, $\mPBH$, and their fractional energy density at the time they form, $\OmegaPBHf$. In \Sec{eq:Dynamical:Constraints}, we first derive the condition on $\OmegaPBHf$ for PBHs to dominate the energy budget of the universe before evaporating, and in \Sec{sec:Constraint:Backreaction}, we identify the region in parameter space where the induced gravitational waves are too abundant and lead to a backreaction problem.
\subsection{Conditions for a PBHs dominated phase}
\label{eq:Dynamical:Constraints}
The mass of a primordial black hole corresponds to some fraction $\xi$ of the mass contained inside a Hubble volume at the time of formation, $\mPBH= 4 \pi \gamma \rho_\mathrm{f}H^{-3}_\mathrm{f}/3 $. Making use of Friedmann's equation, $H^2=\rho_\mathrm{tot}/(3 \Mp^2)$, and assuming that $\gamma \sim 1$, this leads to $\mPBH=4\pi\Mp^2/H_\mathrm{f}$. If PBHs form during the radiation era, since they behave as pressureless matter, their relative contribution to the background energy density grows as $\OmegaPBH\propto a$, so they come to dominate the universe content when the scale factor reaches $a_\ud = a_\mathrm{f}/\OmegaPBHf$. In a radiation era, $H\simeq 1/(2t)\propto 1/a^2$, so this happens at a time $t_\ud = \mPBH/(8 \pi \Mp^2 \OmegaPBHf^2)$.

However, PBHs may have evaporated before that time. The Hawking evaporation time of a black hole with mass $\mPBH$ is given by~\cite{Hawking:1974rv}
\bea
\label{t_evap-t_f}
 t_\mathrm{evap} = \frac{160}{\pi g_\ueff}\frac{m^3_\mathrm{PBH}}{\Mp^4}\,,
 \eea
where $g_\ueff$ is the effective number of degrees of freedom. In numerical applications we take $g_\ueff=100$ since it is the order of magnitude predicted by the Standard Model before the electroweak phase transition~\cite{Kolb:1990vq}, but note that it could assume larger values in extensions to the Standard Model and for this reason we keep it generic in the following formulas. Requiring that $t_\mathrm{evap}>t_\ud$ leads then to the condition
\bea
\label{eq:domain:OmegaPBHf}
\OmegaPBHf >   10^{-15} \sqrt{\frac{g_\ueff}{100}} \frac{10^9\mathrm{g}}{\mPBH}\, .
\eea
As already stressed, one must also impose that PBHs evaporate before big-bang nucleosynthesis takes place, \ie that $H_\mathrm{evap} \simeq 1/(2 t_\mathrm{evap}) > H_\mathrm{BBN}=\sqrt{\rho_\mathrm{BBN}/(3 \Mp^2)}$. With $\rho^{1/4}_\mathrm{BBN}\sim 1 \mathrm{MeV}$, this leads to $ \mPBH< 10^9 \mathrm{g} $ as already mentioned in \Sec{sec:intro}. Note that since PBHs form after inflation, one must also ensure that $H_\mathrm{f}<H_\mathrm{inf}$. In single-field slow-roll models of inflation, the current upper bound on the tensor-to-scalar ratio~\cite{Akrami:2018odb} imposes that $\rho_\mathrm{inf}^{1/4}\lesssim 10^{16}\, \GeV$, and this leads to $\mPBH>10 \mathrm{g}$, so the relevant range of PBH masses is given by
\bea
\label{eq:domain:mPBHf}
10 \mathrm{g}< \mPBH< 10^9 \mathrm{g}\, .
\eea
The relations~\eqref{eq:domain:OmegaPBHf} and~\eqref{eq:domain:mPBHf} define the domain in parameter space where to carry out our calculation.
\subsection{Avoiding the gravitational-wave backreaction problem}
\label{sec:Constraint:Backreaction}
Let us recall that in the PBH-dominated era, the power spectrum of the Bardeen potential is given by \Eq{eq:PowerSpectrum:Phi:PBHdom}, where the UV-cutoff wavenumber $k_\mathrm{UV}$ was defined in \Eq{k_UV}.
Making use of the relation $\bar{r}=\left(\frac{3\mPBH}{4\pi \rhobarPBH}\right)^{1/3}$ given below \Eq{eq:2pt:function:real:space:main:text}, and since, as explained at the beginning of \Sec{eq:Dynamical:Constraints}, $\mPBH=4\pi\Mp^2/H_\mathrm{f}$, it can be expressed as
\bea
k_\mathrm{UV}=\calH_\mathrm{f}\OmegaPBHf^{1/3}\, .
\eea
The tensor power spectrum can then be obtained by plugging \Eq{eq:PowerSpectrum:Phi:PBHdom} into \Eq{Tensor Power Spectrum}, and the power spectrum of the energy density contained in gravitational waves, given in \Eq{Omega_GW}, takes the form 
\bea
\label{Omega_2nd_order_RD_newtonian_gauge_UV_cutoff}
 \OmegaGW (\eta,k)  = \frac{4}{75\pi^2}\left(\frac{k}{aH}\right)^{2}\left(\frac{k}{k_\mathrm{UV}}\right)^6  \, 
{\cal F}\left(\frac{k}{a_{\rm d}H_{\rm d}},  \OmegaPBHf \right)
\eea
with 
\bea
\label{curly F}
{\cal F}(y, \OmegaPBHf )=\int_{0}^{\Lambda} \mathrm{d}v\int_{|1-v|}^{\min(\Lambda,1+v)}\mathrm{d}u\left[ \frac{4v^2 - (1+v^2-u^2)^2}{4\left(3+\frac{4}{15}y^2v^2\right)\left(3+\frac{4}{15}y^2u^2\right)}\right]^{2} {uv}\, .
\eea
Here, we have used that $\overline{I^{2}}=100/9$ in a matter-dominated era and in the sub-Hubble limit, \ie $k\gg \calH$, as shown in \App{rho_GW}.\footnote{Hereafter we restrict the calculation of the power spectrum of the energy density contained in gravitational wave to sub-Hubble scales, since, as mentioned in \Sec{sec:Stress:Energy:GW}, only for those scales is the interpretation of the energy density unambiguous. However, as will be made clear below, the integrated energy density carries little dependence on the lowest wavenumber one considers, which makes our results independent of the infrared cutoff.\label{footnote:IRcutoff}}
In the above expression, we have introduced $y=k/\calH_\mathrm{d}$, and  the upper bound  of the integral over $v$ is given by
\bea
\label{Lambda definition}
\Lambda = \frac{k_\mathrm{UV}}{k} = y^{-1}\Omega^{-2/3}_\mathrm{PBH,f}\,.
\eea
As noted above \Eq{eq:OmegaGW:def}, due to energy-momentum conservation, the tensor power spectrum is non vanishing only  at scales $k<2k_\mathrm{UV}$, which implies that $\Lambda>\frac{1}{2}$.
\begin{figure}[t!]
\begin{center}
  \includegraphics[width=0.496\textwidth, clip=true]
                  {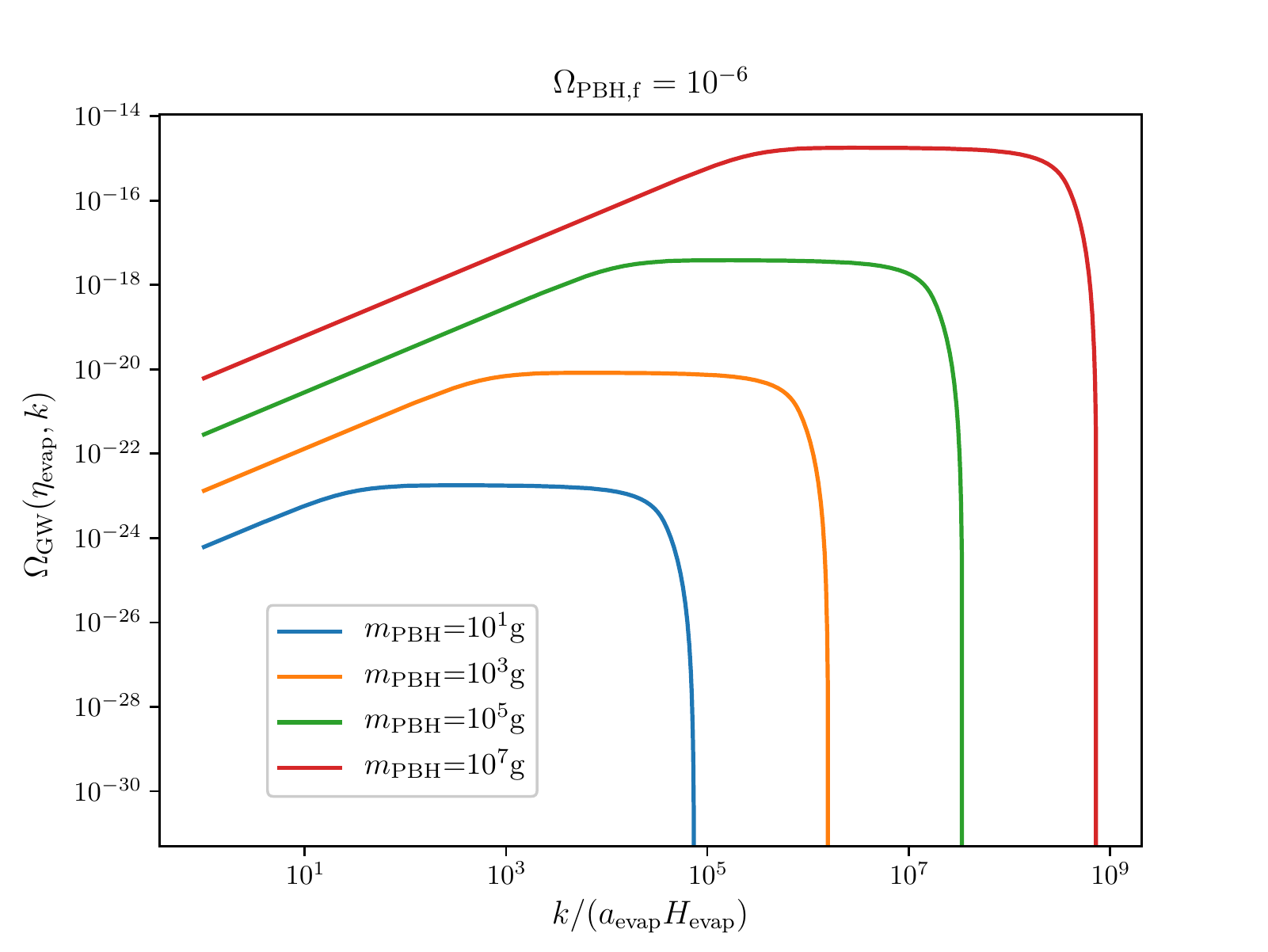}
   \includegraphics[width=0.496\textwidth, clip=true]
                  {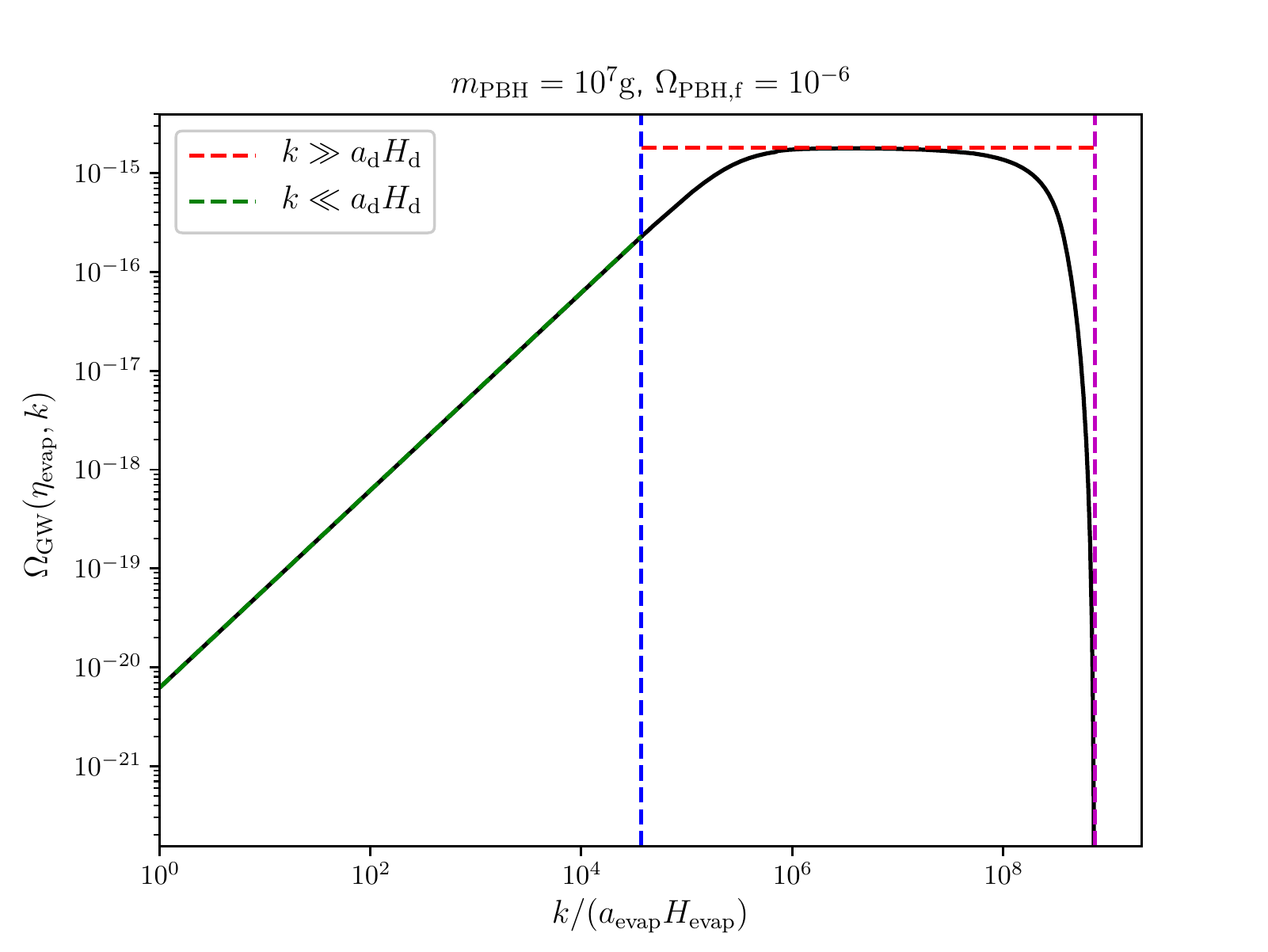}
\caption{
Power spectrum of the energy density contained in gravitational waves as a function of the wavenumber normalised to the comoving Hubble scale at the time of black-hole evaporation. The left panel is for $\OmegaPBHf=10^{-6}$ and shows a few values of the mass $\mPBH$. The right panel focuses on the case $\OmegaPBHf=10^{-6}$ and $\mPBH=10^7 \mathrm{g}$, where the analytical approximations derived in the regimes $k\ll \calH_\ud$ [\Eq{Omega_2nd_order_RD_newtonian_gauge_analytical_approximation - k<<a_dH_d}, shown as the dashed green line] and $k\gg \calH_\ud$ [\Eq{Omega_2nd_order_RD_newtonian_gauge_analytical_approximation - k>>a_dH_d}, shown as the dashed red line] are superimposed. The vertical, dashed blue line stands for $k=\calH_\ud$, \ie for the comoving Hubble scale at the onset time of the PBH-dominated phase, where the power spectrum changes slope. The vertical, dashed magenta line corresponds to $k=2k_\mathrm{UV}$, above which non-linear effects are expected to become important, and this explains why this regime is discarded from our analysis.}
\label{fig:Omega_GW_newtonian_gauge_RD_to_MD}
\end{center}
\end{figure}
The double integral appearing in \Eq{curly F} can be computed numerically, and in the left panel of \Fig{fig:Omega_GW_newtonian_gauge_RD_to_MD}, the result is displayed for $\OmegaPBHf=10^{-6}$ and a few values of $\mPBH$. One can see that the amplitude of the power spectrum increases with the mass $\mPBH$, since larger masses take longer to evaporate and thus give more time for gravitational waves to be produced.

Further analytical insight can be gained by expanding the double integral appearing in \Eq{curly F} in the two  regimes $ y\ll 1$ and  $y\gg 1 $. This is done in detail in \App{app:the:double:integral}, where it is shown that
\bea
\label{eq:calF:approx}
{\cal F}(y, \OmegaPBHf )\simeq 
\begin{cases}
\frac{1125 \sqrt{5} \pi}{256 y^7} \mathrm{\;for\;}y\ll 1 \mathrm{\;and\;}   \OmegaPBHf \ll 1 \\ 
 \frac{50625\pi^2}{2048 y^8}\mathrm{\ for\ }y\gg 1
\end{cases} .
\eea
In the case $y\ll 1$, we only give the limit of the expression where $\OmegaPBHf\ll1 $ since   the value of  $ \OmegaPBHf $  where a backreaction problem occurs will turn out to be much smaller than one. The full expression for an arbitrary value of  $\OmegaPBHf $ can be found in \App{app:the:double:integral}.
Plugging the above results into \Eq{Omega_2nd_order_RD_newtonian_gauge_UV_cutoff}, one obtains
\begin{eqnarray}
\label{Omega_2nd_order_RD_newtonian_gauge_analytical_approximation - k<<a_dH_d}
\OmegaGW(\eta_\mathrm{evap},k\ll \calH_\mathrm{d}) 
 & \simeq &\left(\frac{3\sqrt{5}}{4}\right)^{5/3}\frac{1}{\pi}\left(\frac{g_\mathrm{eff}}{100}\right)^{-2/3} \frac{k}{\calH_\ud}\left(\frac{ \mPBH }{\Mp}\right)^{4/3}\Omega^{16/3}_\mathrm{PBH,f}\,  ,
 \quad\quad
 \\
\label{Omega_2nd_order_RD_newtonian_gauge_analytical_approximation - k>>a_dH_d}
\OmegaGW(\eta_\mathrm{evap},k\gg \calH_\mathrm{d}) 
& \simeq & \frac{135}{64}\left(\frac{45}{2}\right)^{1/3}\left(\frac{g_\mathrm{eff}}{100}\right)^{-2/3} \left(\frac{ \mPBH }{\Mp}\right)^{4/3}\Omega^{16/3}_\mathrm{PBH,f}\, .
\end{eqnarray}
Replacing the prefactors with their numerical values, this gives rise to
\bea
\label{Omega_2nd_order_RD_newtonian_gauge_analytical_approximation}
\OmegaGW(\eta_\mathrm{evap},k)
\simeq
10^{19} \left(\frac{g_\ueff}{100}\right)^{-2/3}\left(\frac{ \mPBH }{10^{9}\mathrm{g}}\right)^{4/3}\Omega^{16/3}_\mathrm{PBH,f}\times
\begin{cases}
\frac{k}{\calH_\ud} \quad \mathrm{for}\quad k\ll \calH_\ud \\
8 \quad \mathrm{for}\quad k\gg \calH_\ud
\end{cases}.
\eea
Those formulas confirm that the amplitude of the power spectrum increases with the mass $\mPBH$, as already noticed in the left panel of \Fig{fig:Omega_GW_newtonian_gauge_RD_to_MD}. They also show that the energy density contained in gravitational waves increases with $\OmegaPBHf$, as one may have expected. The power spectrum is thus made of two branches: a branch scaling as $k$ for $k\ll \calH_\ud$, and a scale-invariant branch for $\calH_\ud \ll k \ll k_{\mathrm{UV}}$. The two approximations~\eqref{Omega_2nd_order_RD_newtonian_gauge_analytical_approximation - k<<a_dH_d} and~\eqref{Omega_2nd_order_RD_newtonian_gauge_analytical_approximation - k>>a_dH_d} are superimposed to the numerical result in the right panel of \Fig{fig:Omega_GW_newtonian_gauge_RD_to_MD}, where one can check that the agreement is indeed good.

Note that as $k$ approaches its maximal value, $2 k_{\mathrm{UV}}$, the approximation fails to describe the sharp cutoff in the power spectrum. This is because, when deriving \Eq{eq:calF:approx} in \App{app:the:double:integral}, we also assumed that $k\ll k_{\mathrm{UV}}$. However, this concerns a small range of modes only, and has little impact on the estimated amount of the overall energy density, as we shall now see.

\begin{figure}[h!]
\begin{center}
  \includegraphics[width=0.796\textwidth,  clip=true]
                  {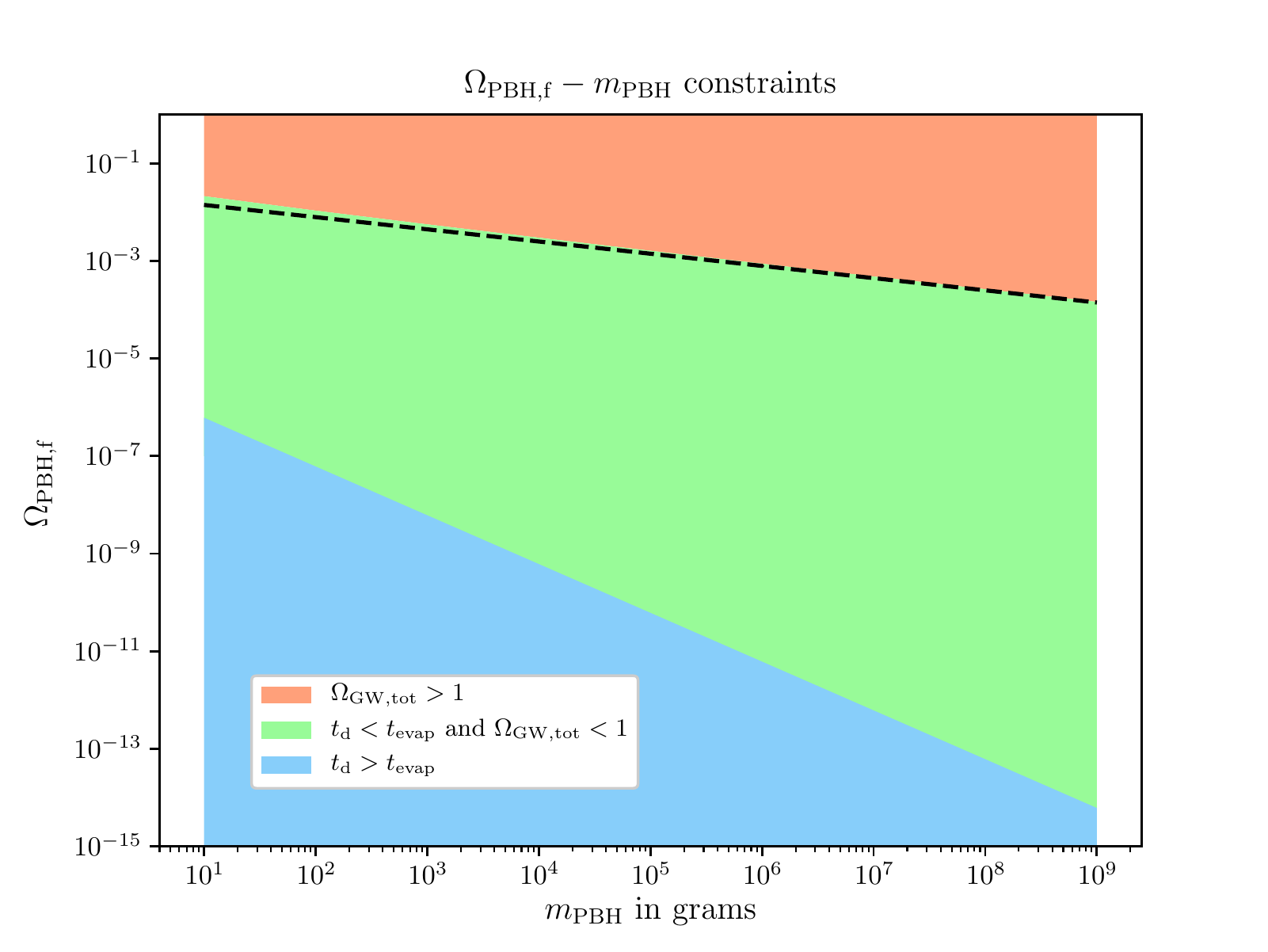}
\caption{Fractional energy density contained in gravitational waves, $\OmegaGWtot$, at the time of PBHs evaporation, as a function of the two parameters of the problem, namely the mass $\mPBH$ of PBHs, and their relative abundance $\OmegaPBHf$ at the time they form. The orange
region corresponds to $\OmegaGW(t_\mathrm{evap})>1$ and leads to a backreaction problem. This region is therefore excluded. The blue region corresponds to values of $\OmegaPBHf$ such that PBHs never dominate the universe, while the green region is such that a transient PBH-dominated phase does take place, but does not lead to a backreaction problem. The dashed black line corresponds to the analytical approximation~\eqref{Omega_f - upper bound in terms of M_0}, which is an analytical estimate of the upper bound on $\OmegaPBHf$ imposed by the need to avoid the backreaction problem. One can check that it indeed provides a good approximation for the boundary between the orange and the green regions. The boundary between the blue and green
regions is given by \Eq{eq:domain:OmegaPBHf}.}
\label{fig:Omega_f_newtonian_gauge_RD_to_MD_constraints}
\end{center}
\end{figure}
The integrated energy density contained in gravitational waves is given by \Eq{eq:OmegaGW:def}, so its fractional contribution to the overall energy budget reads
\beq
\label{eq:OmegaGW:tot}
\OmegaGWtot(\eta) = \int\mathrm{d}\ln k \  \OmegaGW (\eta,k)\, .
\eeq
This integral can be performed numerically, making use of \Eqs{Omega_2nd_order_RD_newtonian_gauge_UV_cutoff} and~\eqref{curly F}, and the result is displayed in \Fig{fig:Omega_f_newtonian_gauge_RD_to_MD_constraints}. When $\OmegaPBHf$ is larger than a certain value, one has $\OmegaGWtot(\eta_\mathrm{evap})>1$, which leads to a backreaction problem. One can therefore derive an upper bound on $\OmegaPBHf$ such that this does not happen, which corresponds to the boundary between the green and the orange region in  \Fig{fig:Omega_f_newtonian_gauge_RD_to_MD_constraints}. An analytical approximation of this upper bound can also be obtained by integrating \Eqs{Omega_2nd_order_RD_newtonian_gauge_analytical_approximation - k<<a_dH_d} and~\eqref{Omega_2nd_order_RD_newtonian_gauge_analytical_approximation - k>>a_dH_d} over $k$,\footnote{Since, for $k\ll \calH_\ud$, $\OmegaGW\propto k$, see \Eq{Omega_2nd_order_RD_newtonian_gauge_analytical_approximation - k<<a_dH_d}, the integral~\eqref{eq:OmegaGW:tot} converges at low $k$, and in the regime where $\calH_\mathrm{evap}\ll \calH_\ud$, one can simply neglect the contribution coming from its lower bound, which justifies the remark made in footnote~\ref{footnote:IRcutoff}.} and \Eq{eq:OmegaGW:tot} leads to
\bea
\OmegaGWtot(\eta_\mathrm{evap})=\mu\left[\kappa-\ln(\OmegaPBHf)\right]\OmegaPBHf^{16/3}\, ,
\eea
with 
\bea
\mu= \left(\frac{45}{2}\right)^{4/3}\frac{1}{16}\left(\frac{g_\ueff}{100}\right)^{-2/3} \left(\frac{\mPBH}{\Mp}\right)^{4/3}
\quad\mathrm{and}\quad
\kappa = \frac{4}{3\sqrt{5}\pi}+\frac{3}{2}\ln(2)\, .
\eea
The equation $\OmegaGWtot=1$ can be solved by means of the Lambert function~\cite{Olver:2010:NHM:1830479:Lambert}, and one obtains
\bea
\OmegaPBHf^\umax=\left[ -\frac{3 \mu}{16}W_{-1}\left(-\frac{16}{3\mu}\ee^{-\frac{16\kappa}{3}}\right)\right]^{-3/16},
\eea
where $W_{-1}$ is the ``$-1$''-branch of the Lambert function. Since $\mPBH>10\mathrm{g}$ (see \Eq{eq:domain:mPBHf}), one has $\mu\gg 1$ while $\kappa$ is of order one, so the argument of the Lambert function is close to zero. In this regime, the Lambert function can be approximated by a logarithmic function. Given the mild dependence of the logarithm on its argument, and since $\mPBH$ varies over 8 orders of magnitude ``only'', in practice, it can be approximated as constant (and evaluated for a central mass in the range, namely $\mPBH=10^{5}\mathrm{g}$), and one obtains 
\bea
\label{Omega_f - upper bound in terms of M_0}
\OmegaPBHf^\umax \simeq 1.4 \times 10^{-4}\left(\frac{10^9\mathrm{g}}{ \mPBH }\right)^{1/4}\,.
\eea 

This approximation is superimposed in \Fig{fig:Omega_f_newtonian_gauge_RD_to_MD_constraints} and one can check that it provides an accurate estimate of the boundary of the region where gravitational waves are over produced (orange region). Recalling that $\mPBH>10\mathrm{g}$, \Eq{Omega_f - upper bound in terms of M_0} excludes the possibility to form PBHs in such an abundant way that they dominate the universe content right upon their formation time (\ie the value $\OmegaPBHf=1$ is excluded). Otherwise, there exists a region (displayed in green in \Fig{fig:Omega_f_newtonian_gauge_RD_to_MD_constraints}) where PBHs happen to dominate the universe content at a later time, but do not lead to a gravitational-wave backreaction problem. Note that our calculation does not apply to the blue region in \Fig{fig:Omega_f_newtonian_gauge_RD_to_MD_constraints}, which is where PBHs never dominate the universe, but it is clear that no gravitational-wave backreaction problem can happen there.
\section{Conclusions}
\label{sec:Conclusions}
In this work, we have studied the gravitational waves induced at second order by the gravitational potential of 
a gas of primordial black holes. In particular, we have considered scenarios where ultralight PBHs, with masses $\mPBH<10^{9}\mathrm{g}$, dominate the universe content during a transient period~\cite{Anantua:2008am, Zagorac:2019ekv, Martin:2019nuw, Inomata:2020lmk}, before Hawking evaporating. Neglecting clustering at formation~\cite{Desjacques:2018wuu, MoradinezhadDizgah:2019wjf}, the Poissonian fluctuations in their number density underlay small-scale density perturbations, which in turn induce the production of gravitational waves at second order.

In practice, we have computed the gravitational-wave energy spectrum, as well as the integrated energy density of gravitational waves, as a function of the two parameters of the problem, namely the mass of the PBHs, $\mPBH$ (assuming that all black holes form with roughly the same mass~\cite{MoradinezhadDizgah:2019wjf}), and their relative abundance at formation $\OmegaPBHf$. This calculation was performed both numerically and by means of well-tested analytical approximations. We have found that the amount of gravitational waves increases with $\mPBH$, since heavier black holes take longer to evaporate, hence dominate the universe for a longer period; and with $\OmegaPBHf$, since more abundant black holes dominate the universe earlier, hence for a longer period too. 
\begin{figure}[h!]
\begin{center}
  \includegraphics[width=0.796\textwidth,  clip=true]
                  {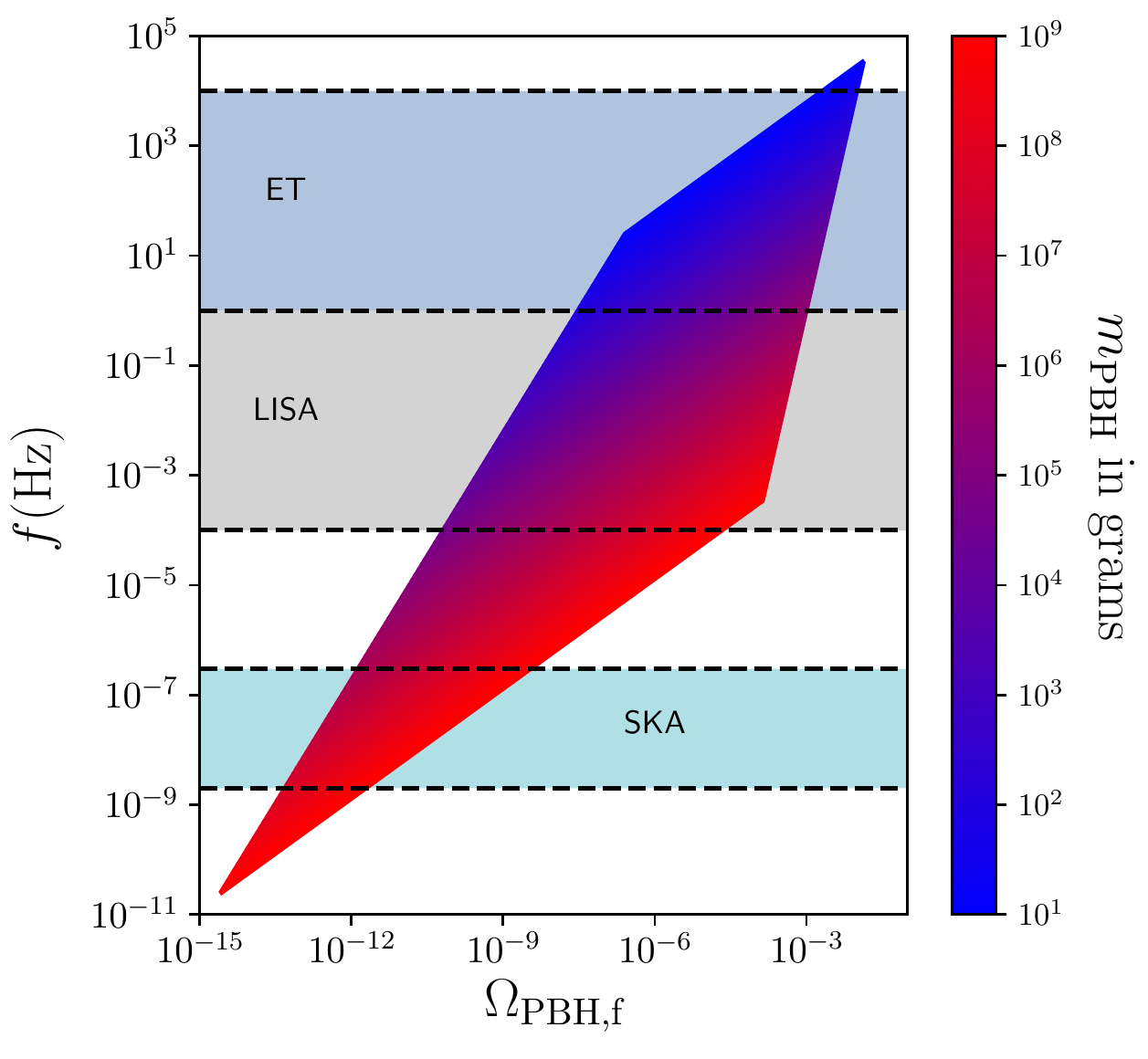}
\caption{Frequency at which the gravitational waves induced by a dominating gas of primordial black holes peak, as a function of  
their energy density fraction at the time they form, $\OmegaPBHf$ (horizontal axis), and their mass $\mPBH$ (colour coding).
The region of parameter space that is displayed corresponds to values of $\mPBH$ and $\OmegaPBHf$ such that black holes dominate the universe content for a transient period, see \Eq{eq:domain:OmegaPBHf}, that they form after inflation and Hawking evaporate before big-bang nucleosynthesis, see \Eq{eq:domain:mPBHf}, and that the induced gravitational waves do not lead to a backreaction problem, see \Eq{Omega_f constraints}. In practice, \Eq{GW frequency} is displayed with $g_\ueff=100$, $z_\mathrm{eq}=3387$ and $H_0=70\,\mathrm{km}\,\mathrm{s}^{-1}\,\Mpc^{-1}$. For comparison, the detection bands of ET, LISA and SKA are also shown.}
\label{fig:GW frequency}
\end{center}
\end{figure}

Requiring that the energy contained in gravitational waves never overtakes the one of the background universe led us to the constraint
\bea
 \label{Omega_f constraints}
\OmegaPBHf < 1.4\times 10^{-4}\left(\frac{10^9\mathrm{g}}{ \mPBH }\right)^{1/4} .
 \eea
Let us stress that since PBHs with masses smaller than $10^9\mathrm{g}$ evaporate before big-bang nucleosynthesis, they cannot be directly constrained (at least without making further assumption, see \eg \Refa{Dai:2019eei}). To our knowledge, the above constraint is therefore the first one ever derived on ultra-light PBHs. In particular, it shows that scenarios where PBHs dominate from their formation time on, $\OmegaPBHf\simeq 1$, are excluded (given that $m>10\mathrm{g}$ for inflation to proceed at less than $10^{16}\mathrm{GeV}$). 

A word of caution is in order regarding the assumptions underlying the present calculation. Since we have made use of cosmological perturbation theory to assess the amount of induced gravitational waves, we only considered scales where the scalar fluctuations underlain by PBHs remain perturbative, which is why we have imposed an ultra-violet cutoff $k_\mathrm{UV}$ at the scale corresponding to the mean separation distance between PBHs. As shown in \Sec{sec:CurvaturePowerSpectrum}, the maximal value of the gravitational-potential power spectrum in the PBH-dominated phase is thus of order $(\calH_\mathrm{d}/k_\mathrm{UV})^3 =\OmegaPBHf^2$, which is indeed much smaller than one. This confirms that from the point of view of the gravitational potential $\Phi$, or from the point of view of the curvature perturbation $\zeta$, all scalar fluctuations incorporated in the calculation lie in the perturbative regime, which is what is required since the induced gravitational waves are sourced by $\Phi$. However, as also noticed in \Sec{sec:CurvaturePowerSpectrum}, the density contrast $\delta$ associated with the gas of PBHs grows like the scale factor during the PBH-dominated era, contrary to $\zeta$ or $\Phi$ which remain constant. Therefore, there are scales for which $\delta$ grows larger than one during the PBH-dominated era, although $\Phi$ remains much smaller than one. The status of these scales is unclear: the growth of $\delta$ above one may signal the onset of PBH clustering, which might result in the enhancement of the power spectrum above the Poissonian value, which might in turn be responsible for an even larger signal than the one we have computed. In this sense, the bounds we have derived could be conservative only, although a more thorough investigation of the virialisation dynamics at small scales would be required.

Let us also note that we did not account for PBH accretion of the surrounding radiation, which could potentially prolong the PBH lifetime beyond the evaporation time. However, for Bondi-Hoyle type accretion~\cite{1944MNRAS.104..273B}, the PBH accretion rate, $\dot{m}_\mathrm{PBH}$, is proportional to the square of the PBH mass, i.e. $\dot{m}_\mathrm{PBH}\propto m^2_\mathrm{PBH}$. It is therefore less relevant for smaller black holes, and recent analyses~\cite{DeLuca:2020bjf,DeLuca:2020fpg} find that accretion is negligible when $m_\mathrm{PBH}<O(10)M_\odot$. This is the case of the ultralight black holes considered here, which have masses smaller than $10^{9}\mathrm{g}$.

Finally, we should stress out that the condition~\eqref{Omega_f constraints} simply comes from avoiding a backreaction problem, and does not implement observational constraints. However, even if the condition~\eqref{Omega_f constraints} is satisfied, gravitational waves induced by a dominating gas of PBHs might still be detectable in the future with gravitational-waves experiments. Although an accurate assessment of the signal would require to properly resolve the dynamics of the induced gravitational waves during the gradual transition between the PBH-dominated and the radiation-dominated era ~\cite{Inomata:2019zqy}, let us note that since we have found that the energy spectrum peaks at the Hubble scale at the time black holes start dominating, this corresponds to a frequency $f=\calH_\ud/(2\pi a_0)$, where $a_\mathrm{0}$ is the value of the scale factor today and $\calH_\ud$ is the comoving Hubble scale at domination time. This leads to
\bea
\label{GW frequency}
\frac{f}{\mathrm{Hz}} \simeq \frac{1}{\left(1+z_\mathrm{eq}\right)^{1/4}}\left(\frac{H_0}{70\mathrm{kms^{-1}Mpc^{-1}}}\right)^{1/2}\left(\frac{g_\ueff}{100}\right)^{1/6}\OmegaPBHf ^{2/3}\left(\frac{ \mPBH }{10^9\mathrm{g}}\right)^{-5/6} ,
\eea
where $H_0$ is the value of the Hubble parameter today and $z_\mathrm{eq}$ is the redshift at matter-radiation equality. In \Fig{fig:GW frequency}, this frequency is shown in the region of parameter space that satisfies the condition~\eqref{Omega_f constraints}. Covering $14$ orders of magnitude, one can see that it intersects the detection bands of the Einstein Telescope (ET)~\cite{Maggiore:2019uih}, the Laser Interferometer Space Antenna (LISA)~\cite{Audley:2017drz,Caprini:2015zlo} and  the Square Kilometre Array (SKA) facility~\cite{Janssen:2014dka}. This may help to further constrain ultra-light primordial black holes, and set potential targets for these experiments.
\begin{acknowledgments}
We thank Valerio De Luca, Gabriele Franciolini, Keisuke Inomata and Antonio Riotto for useful discussions.
T.~P. acknowledges support from the \textit{Fondation CFM pour la
Recherche} and from the \textit{Onassis Foundation} through the scholarship FZO 059-1/2018-2019.
\end{acknowledgments}
\appendix
\section{Density power spectrum for a Poissonian gas of PBHs}
\label{app:Poisson Statistics}
Let us consider a gas of $N$ PBHs, each of them with the same mass $\mPBH$, and randomly distributed inside a volume $V$. The location of each PBH is random and follows a uniform distribution across the entire volume. We assume that it is not correlated with the location of other PBHs within the gas, which implies that we consider each black hole as a point-like particle, since we neglect the existence of an exclusion zone around the position of the centre of a black hole. This means that length scales smaller than the Schwarzschild radius are not properly described in this setup, which only applies to derive the density field on scales larger than the black holes size.

Let us now consider a sphere of radius $r$ (and of volume $v=4\pi r^3/3$) within the volume $V$, and denote by $P_{n}(r)$ the probability that $n$ PBHs are located inside this volume. For each PBH, the probability to be inside the sphere is given by $v/V$, and the probability to be outside is given by $(V-v)/V$, so one has
\bea
P_{n}(r) & = \binom{N}{n} \left(\frac{v}{V}\right)^n \left(1-\frac{v}{V}\right)^{N-n}\, .
\eea
By denoting $\bar{r}$ the mean distance between black holes, such that $V=4\pi \bar{r}^3 N/3$, this can be written as
\bea
\label{PDF for n PBHS inside r}
P_{n}(r) & =\binom{N}{n}\left(\frac{r^3}{N\bar{r}^3}\right)^{n}\left(1-\frac{ r^3}{N\bar{r}^3}\right)^{N-n}  
\underset{N\to \infty}{\longrightarrow}
\left(\frac{r}{\bar{r}}\right)^{3n}\frac{e^{-\frac{r^3}{\bar{r}^3}}}{n!}
\eea
where we have taken the large-volume limit. Such statistics are referred to as Poissonian. 

The total mass of the PBHs contained within the volume $v$ is given by $n \mPBH$, so the mean PBH energy density within the volume can be written as 
\bea
\rhobarPBH(r) = \frac{n \mPBH}{\frac{4}{3}\pi r^3}\, .  
\eea
By making use of \Eq{PDF for n PBHS inside r}, one can compute the two first moments of this quantity. One first has
\bea
\label{eq:app:rho_bar}
\left\langle \rhobarPBH(r) \right\rangle = \sum_{n=0}^\infty P_n(r) \frac{n \mPBH}{\frac{4}{3}\pi r^3} = \frac{\mPBH}{\frac{4}{3}\pi \bar{r}^3},
\eea
which is independent of $r$ and simply corresponds to the average energy density. One then finds
\bea
\left\langle\rhobarPBH^2(r) \right\rangle = \sum_{n=0}^\infty P_n(r) \left(\frac{n \mPBH}{\frac{4}{3}\pi r^3}\right)^2 =\frac{9\mPBH^2}{16\pi^2 r^6}\left[\left(\frac{r}{
\bar{r}}\right)^{3} + \left(\frac{r}{\bar{r}}\right)^{6}\right]\, .
\eea
Combining the two above results, one obtains the variance of the energy density fluctuation,
\bea
\label{eq:rho2:1}
\left\langle \delta \rhobarPBH^2(r) \right\rangle =\left\langle \rhobarPBH^2(r) \right\rangle-\left\langle \rhobarPBH(r) \right\rangle^2=
\frac{9\mPBH^2}{16\pi^2 \bar{r}^6}\left(\frac{\bar{r}}{r}\right)^{3}\, .
\eea

Let us now describe the gas of PBHs in terms of a fluid with energy density $ \rhoPBH (\bm{x})$, and density contrast $\delta \rhoPBH (\bm{x})/\rho_\mathrm{tot}$, where $\delta \rhoPBH (\bm{x})= \rhoPBH (\bm{x}) - \left\langle \rhobarPBH\right\rangle$ and $\rho_\mathrm{tot}$ is the mean total energy density (comprising PBHs but also other possible fluids). The mean energy density with the volume $v$ can be written as
\bea
\rhobarPBH(r) &=\frac{1}{\frac{4}{3}\pi r^3} \int_{\vert \bm{x} \vert <r} \dd^3\bm{x}   \rhoPBH (\bm{x})\\
&=\left\langle \rhobarPBH \right\rangle + \frac{\rho_\mathrm{tot}}{\frac{4}{3}\pi r^3} \int_{\vert \bm{x} \vert <r} \dd^3\bm{x}   \frac{\delta  \rhoPBH (\bm{x})}{\rho_\mathrm{tot}} \, .
\label{eq:delta:rho:fluid}
\eea
As explained above, the gas of PBHs being Poissonian, the existence of a PBH at location $\bm{x}$ is uncorrelated with the position of a PBH at location $\bm{x}'$, which means that 
\bea
\label{eq:2pt:function:space:ansatz}
\left\langle \frac{\delta  \rhoPBH (\bm{x})}{\rho_\mathrm{tot}}\frac{\delta  \rhoPBH (\bm{x}')}{\rho_\mathrm{tot}}\right\rangle = \xi\,  \delta(\bm{x}-\bm{x}')\, ,
\eea
where  $\xi$ a priori depends on $\bm{x}$ and $\bm{x}'$ (because of statistical homogeneity and isotropy, only through $\vert \bm{x}-\bm{x}'\vert$).
By averaging the square of \Eq{eq:delta:rho:fluid}, one thus obtains
\bea
\label{eq:rho2:2}
\left\langle \delta \rhobarPBH^2(r) \right\rangle =\frac{9\rho_\mathrm{tot}^2}{16\pi^2r^6} \frac{4}{3}\pi r^3\, \xi\, .
\eea
By identifying \Eqs{eq:rho2:1} and~\eqref{eq:rho2:2}, one can read off 
$\xi=3\mPBH^2/(4 \pi\rho_\mathrm{tot}^2 \bar{r}^3)$. 
By introducing the PBH fractional energy density $ \OmegaPBH =\left\langle \rhobarPBH \right\rangle/\rho_{\mathrm{tot}}$, \Eq{eq:2pt:function:space:ansatz} can thus be written as 
\bea
\label{eq:2pt:function:real:space}
\left\langle \frac{\delta  \rhoPBH (\bm{x})}{\rho_\mathrm{tot}}\frac{\delta  \rhoPBH (\bm{x}')}{\rho_\mathrm{tot}}\right\rangle = \frac{4}{3}\pi \bar{r}^3  \OmegaPBH ^2\delta(\bm{x}-\bm{x}')\, ,
\eea
which is the expression we use in \Sec{sec:PBH Distribution}. 
\section{Kinetic and gradient contributions to the gravitational-waves energy}
\label{rho_GW}
In this appendix, we compute the energy density contained in gravitational waves, given in \Eq{rho_GW effective}, which is made of a kinetic contribution and a gradient contribution.
\subsection{Kinetic contribution}
From \Eq{rho_GW effective}, the kinetic contribution to the gravitational-waves energy density is given by
\bea
\label{rho_GW kinetic 1}
\rhoGW^\mathrm{kin}(\eta,\boldmathsymbol{x}) 
& = \frac{\Mp^2}{32 a^2} \sum_{s=+,\times} \overline{ \left\langle{h}^{s,\prime}_{ij}{h}^{s,ij,*,\prime} \right\rangle }
 \\ 
& =  \sum_{s=+,\times} \frac{\Mp^2}{32a^2 \left(2\pi\right)^{3}} \int \dd^3\boldmathsymbol{k}_1 \int \dd^3\boldmathsymbol{k}_2 \overline{ \left\langle h^{s,\prime}_{k_1}h^{s,*,\prime}_{k_2}  \right\rangle} e^{i(\boldmathsymbol{k}_1-\boldmathsymbol{k}_2)\cdot\boldmathsymbol{x}}\, .
\eea
Recall that this expression is valid on sub-Hubble scales only, as discussed in \Sec{sec:Stress:Energy:GW}, and the bar denotes averaging over the oscillations of the tensor fields at those scales.

The time derivative of the tensor mode-function can be computed from differentiating \Eq{tensor mode function}, and one obtains
\bea
\label{hprime_k}
h^{s,\prime}_k(\eta) &= -\mathcal{H}h^s_k(\eta) + 4G_k(\eta,\eta)S^s_k(\eta) + \frac{4}{a(\eta)}\int_{\eta_\mathrm{d}}^{\eta}a(\bar{\eta})G^\prime_k(\eta,\bar{\eta})S^s_k(\bar{\eta})\mathrm{d}\bar{\eta} 
\\ & = -\mathcal{H}h^s_k(\eta) + \frac{4}{a(\eta)}\int_{\eta_\mathrm{d}}^{\eta}a(\bar{\eta})G^\prime_k(\eta,\bar{\eta})S^s_k(\bar{\eta})\mathrm{d}\bar{\eta}\, ,
\eea
where we have used the fact that, as mentioned below \Eq{Green function equation}, $G_k(\eta,\eta)=0$. Hereafter, $G_k'(\eta,\bar{\eta})$ denotes the derivative of $G_k$ with respect to its first argument, $\eta$. On sub-Hubble scales, $k\gg \mathcal{H}$, \Eq{Green function equation} reduces to $G_k''+k^2G_k\simeq 0$, hence $G_k'\sim \pm i k G_k$, and the second term in the last line of \Eq{hprime_k} is of order $k\, h^s_k$ according to \Eq{tensor mode function}, and thus dominates over the first term, i.e. 
\bea
h^{s,\prime}_k(\eta) \simeq \frac{4}{a(\eta)}\int_{\eta_\mathrm{d}}^{\eta}a(\bar{\eta})G^\prime_k(\eta,\bar{\eta})S^s_k(\bar{\eta})\mathrm{d}\bar{\eta}\, .
\eea
The two-point function of $h^{s,\prime}_k$ can then be computed from the source correlator in exactly the same way the two-point function of $h^{s}_k$ was evaluated for the power spectrum in \Sec{sec:tensor:power:spectrum}, and one obtains
\bea
\label{h_prime correlator}
\langle h^{r,\prime}_{k_1}h^{s,*,\prime}_{k_2}\rangle & =\delta^{(3)}(\boldmathsymbol{k}_1 - \boldmathsymbol{k}_2) \delta^{rs} \frac{2\pi^2}{k^3_1}\mathcal{P}_{h^\prime}(\eta,k_1)
\eea
where
\bea
\mathcal{P}_{h^\prime}(\eta,k) = 
4 k^2\int_{0}^{\infty} \mathrm{d}v\int_{|1-v|}^{1+v}\mathrm{d}u \left[ \frac{4v^2 - (1+v^2-u^2)^2}{4uv}\right]^{2}{J^2(u,v,x)}\mathcal{P}_\Phi(kv)\mathcal{P}_\Phi(ku) 
\eea
and
\bea
\label{I_kin function}
J(u,v,x) = \int_{x_\mathrm{d} }^{x} \mathrm{d}\bar{x}\frac{a(\bar{\eta})}{a(\eta)}G^\prime_{k}(\eta,\bar{\eta})F_k(u,v,\bar{x}).
\eea
Combining these results together, \Eq{rho_GW kinetic 1} gives rise to
\bea
\label{rho_GW kinetic 2}
\rhoGW^\mathrm{kin}(\eta,\boldmathsymbol{x}) & = 
 \int \mathrm{d}\ln k \frac{\mathrm{d}\rho^\mathrm{kin}_\mathrm{GW}(k)}{\mathrm{d}\ln k}
\eea
with
\bea
\frac{\mathrm{d}\rhoGW^\mathrm{kin}(k)}{\mathrm{d}\ln k} = \frac{\Mp^2}{16a^2} \overline{\mathcal{P}}_{h^\prime}(\eta,k) ,
\eea
so the fractional energy density contained in gravitational waves can be written as 
\bea
\OmegaGW^\mathrm{kin}\left(\eta,k\right) = \frac{\overline{\mathcal{P}}_{h'} (\eta,k)}{48 a^2(\eta) H^2(\eta)}.
\eea
\subsection{Gradient contribution}
The gradient contribution to the energy density contained in gravitational waves can be derived in the same way, and the result is presented in \Sec{sec:tensor:power:spectrum}. The fractional gradient energy density is given by \Eq{Omega_GW}, so the total fractional energy density reads
\bea
\label{Omega_GW effective}
 \OmegaGW (\eta,k) = \Omega^\mathrm{kin}_\mathrm{GW}(\eta,k)  + \Omega^\mathrm{grad}_\mathrm{GW}(\eta,k)  =  \frac{k^2}{48 a^2(\eta) H^2(\eta)}\left[\frac{\overline{\mathcal{P}}_{h'}(\eta,k)}{k^2}+ \overline{\mathcal{P}}_{h}(\eta,k)\right] .
\eea
\subsection{Gravitational-wave energy in a matter dominated era}
As explained in the main text, our goal is to compute the energy density contained in gravitational waves at the time where PBHs evaporate, \ie at the end of the PBH-dominated epoch. Since PBHs drive a pressureless matter-dominated phase, we now specify the above formulas to such an epoch. As explained in \Sec{sec:2nd:order:GWs}, in a matter era, the Bardeen potential is, up to a decaying mode, constant in time, hence $T_\Phi(x)= 1$. From \Eq{f function - w =0}, one then has 
$F=10/3$. 
By specifying \Eq{Green function analytical} to the case where the equation-of-state parameter vanishes, $w=0$, so $\nu =\frac{3}{2}$, one has 
\bea
\label{Green function analytical MD}
kG_k(\eta,\bar{\eta}) = 
\frac{1}{x\bar{x}}\left[ (1+x\bar{x})\sin(x-\bar{x}) - (x-\bar{x})\cos(x-\bar{x})\right] .
\eea
This allows one to compute the $I$ integral, defined in \Eq{I function}, exactly, and one finds 
\bea
 \label{I_MD with x_d not zero} 
I^2(x)  & = \frac{100}{9}\left[ 1 + \cos(x - x_\mathrm{d}) \left(\frac{3}{x^2}-\frac{3x_\ud}{x^3}-\frac{x_\ud^2}{x^2}\right)
- \sin(x - x_\mathrm{d}) \left(\frac{3}{x^3}+\frac{3 x_\ud}{x^2}-\frac{x_\ud^2}{x^3}\right)
\right]^2 .
\eea
In the sub-Hubble limit, $x\gg 1$ and this reduces to $I^2 \simeq 100/9$. The procedure of averaging over the oscillations becomes trivial in this limit (since there are none) and one simply has $\overline{I^2}\simeq 100/9$.
Similarly, the $J$ integral, defined in \Eq{I_kin function}, can be performed exactly,
\bea
 \label{I_kin with x_d not zero} 
J^2(x)   =  & \frac{100}{9} \biggl[\frac{2}{x} 
-\cos (x-x_\mathrm{d}) \left(\frac{3}{x^3}-\frac{3 x_\ud}{x^4}+\frac{3 x_\ud}{x^2}-\frac{x_\ud^2}{x^3}\right)
\nonumber \\ & 
+\sin (x-x_\mathrm{d}) \left(\frac{3}{x^4}-\frac{3}{x^2}+\frac{3 x_\ud}{x^3}-\frac{x_\ud^2}{x^4}+\frac{x_\ud^2}{x^2}\right)
 \biggr]^2,
\eea
which reduces to $J^2 = \overline{J^2}\simeq 4  \overline{I^2}/x^2$ in the sub-Hubble limit.

As a consequence, in \Eq{Omega_GW effective}, the kinetic term is suppressed by a factor $x^{-2}\ll 1$ compared to the gradient term, which justifies the statement made in \Sec{sec:Stress:Energy:GW} that the gradient term provides the main contribution. This can be understood from the presence of the source term in \Eq{Tensor Eq. of Motion}, which, in a matter-dominated era where $T_\Phi(x)=1$, becomes constant in time [see \Eq{eq:Source:def}]. In the sub-Hubble limit, this forces the mode function $h^{s}_\boldmathsymbol{k}$ towards a constant solution $h^{s}_\boldmathsymbol{k}\simeq \frac{4 S^s_\boldmathsymbol{k}}{k^2}$, which therefore carries little kinetic energy.
\section{Approximation for the double integral in $\OmegaGW$}
\label{app:the:double:integral}
In this appendix, we expand the double integral appearing in \Eq{curly F}, ${\cal{F}}(y, \OmegaPBHf)$, in the two limits $y\ll 1$ and $y\gg 1$. For explicitness, let us introduce the function 
\bea
h(u,v,y) \equiv \left[\frac{4v^2 - (1+v^2-u^2)^2}{4uv}\right]^{2}u^3v^3 \left(3+\frac{4}{15}y^2v^2\right)^{-2}\left(3+\frac{4}{15}y^2u^2\right)^{-2},
\eea
in terms of which \Eq{curly F} can be written as 
\bea
{\cal{F}}(y, \OmegaPBHf) =  \int_{0}^{\Lambda} \mathrm{d}v\int_{|1-v|}^{\min(\Lambda,1+v)}\mathrm{d}u\, h(u,v,y),
\eea
where we recall that $\Lambda = y^{-1}\OmegaPBHf^{-2/3}$, see \Eq{Lambda definition}. A primitive of the function $h(u,v,y)$ with respect to the $u$ variable is given by
\bea
H(u,v, y) & = \frac{75 v}{32768 y^{10}\left(3 + \frac{4}{15}v^2 y^2\right)^2}\biggl\{64\left(u^2 - v^2-1\right)^3 y^6  \\ & - 48 \left(1 - u^2 + v^2\right)^2 y^4 \left[45 + 4(1 + v^2) y^2\right] \\ &
- \frac{3 \left[2025 + 360 (1 + v^2) y^2 + 16(v^2-1)^2 y^4\right]^2}{45 + 4u^2 y^2}  \\ &  
 + 12 (u^2 - v^2 -1) y^2 \left[ 6075 +  1080 (1 + v^2) y^2 + 16(3 - 2 v^2 + 3 v^4) y^4\right] \\ &
- 12 \biggl[ 91125 +   24300 (1 + v^2) y^2 + 720(3 + 2 v^2 + 3 v^4) y^4   \\ & + 64( v^2-1)^2 (1 + v^2) y^6\biggr]\ln\left(45 + 4 u^2 y^2\right) \biggr\} \, .
\eea
 One can indeed check that $ \frac{\partial H(u,v,y)}{\partial u} = h(u,v,y)$. Our next step is to split the remaining integral over $v$ at the splitting points $v=1$ and $v=\Lambda-1$, according to
\bea
\label{eq:double:integral:splitting}
{\cal F}(y, \OmegaPBHf) & =
\int_{0}^{1} \left[ H(1+v, v, y) - H(1-v,v,y) \right] \mathrm{d} v \\ & + \int_{1}^{\Lambda  - 1}\left[ H(1+v, v, y) - H(v-1,v,y) \right] \mathrm{d} v \\ & + \int_{\Lambda - 1}^{\Lambda }\left[ H(\Lambda, v, y) - H(v-1,v,y) \right] \mathrm{d} v 
  \\ &  \equiv
 {\cal F}_1(y) + {\cal F}_2(y,\OmegaPBHf) + {\cal F}_3(y,\OmegaPBHf),
\eea
which defines the three integrals ${\cal F}_1(y)$, ${\cal F}_2(y,\OmegaPBHf)$ and ${\cal F}_3(y,\OmegaPBHf)$. Hereafter, we will not try to resolve the shape of $\OmegaGW$ as one approaches the UV cutoff scale, so we will work under the condition $k\ll k_\mathrm{UV}$, hence $\Lambda\gg 1$.
\subsection{The $y\ll 1$ regime}
Let us first consider the regime where $y\ll 1$. In the first integral $\mathcal{F}_1(y)$, 
it can be shown that 
the integrand is maximal when $v$ is close to one, hence $y$ is the only small parameter of the problem. When expanding the integrand in $y$, one obtains a constant value at leading order, $ H(1+v, v, y) - H(1-v,v,y) \propto y^0$, hence the integral features quantities of order one only, and one has
\bea
\mathcal{F}_1(y) = \order{1}.
\eea 
In the second integral, $\mathcal{F}_2(y)$, the integrand is maximal when $v$ is of order $1/y$. It is therefore convenient to perform the change of integration variables  $t=yv$, such that the integrand is maximal when $t$ is of order one, hence $y$ is again the only small parameter, in terms of which the integrand can be expanded,
\bea
\label{eq:y<<1:F2}
{\cal F}_2(y,\OmegaPBHf) &= \int_{y}^{y(\Lambda  - 1)}\left[ H\left(1+\frac{t}{y}, \frac{t}{y}, y\right) - H\left(\frac{t}{y}-1,\frac{t}{y},y\right) \right] \frac{\mathrm{d} t}{y} \\
& \simeq \frac{54000}{ y^7 }  \int_{y}^{y(\Lambda  - 1)}  \frac{t^6}{\left(45+4t^2\right)^4}\, .
\eea
Similarly, in the third integral, since $v$ is of order $\Lambda$, one can perform the change of integration variable $t = v - \Lambda +1$, such that $0\leq t\leq 1$. Upon expanding in $y$, one then finds that 
\bea
{\cal F}_3(y,\OmegaPBHf) &\propto \frac{\OmegaPBHf^{4/3}}{y^6}\, .
\eea
In the limit where $y\ll 1$, since $\OmegaPBHf<1$, the integral ${\cal F}_2(y,\OmegaPBHf)$ is therefore the dominant one. Noting that the remaining integral over $t$ in \Eq{eq:y<<1:F2} can be performed exactly, this leads to
\bea
{\cal F}(y\ll 1,\OmegaPBHf) \simeq 
\frac{1125}{128y^{7}} & \left[\sqrt{5}\mathrm{arctan}\left(\frac{2}{3\sqrt{5} \OmegaPBHf^{2/3}}\right)
\right. \\ & \quad\left.
-6 \frac{176 \OmegaPBHf^{2/3}+2400 \OmegaPBHf^{2}+10125 \OmegaPBHf^{10/3}}{\left(4+45\OmegaPBHf^{4/3}\right)^3}\right] .
\eea
\subsection{The $y\gg 1$ regime}
Similar techniques can be employed to study the regime where $y\gg 1$. In the first integral, the integrand is maximal when $v$ is close to one, so $1/y$ is the only small parameter, and expanding in $1/y$ leads to
\bea
\mathcal{F}_1(y) & \simeq  \frac{16875}{2048 y^8} \int_0^1
\frac{\dd v}{ v^3 }\left[ 4 v (3 - 2 v^2 + 3 v^4)  + 6 \left(v^2-1\right)^2 (v^2+1)\ln\left( \frac{1-v}{1+v}\right)\right]\\ &
= \frac{16875}{4096}\frac{3\pi^2 - 16}{y^8}\, .
\eea
In the second integral, the integrand is maximal at values of $v$ of order one again, so one can expand in $1/y$ and obtain
\bea
\mathcal{F}_2(y,\OmegaPBHf)  & \simeq 
\frac{16875}{2048 y^8}
 \int_{1}^{\Lambda -1} \frac{\dd v}{v^3}
 \left[
 4v(3-2v^2+3v^4)
 + 6 \left(v^2-1\right)^2 (v^2+1)\ln\left( \frac{v-1}{1+v}\right) \right]\\ &
= \frac{16875}{4096}\frac{3\pi^2 + 16}{y^8}
\, ,
\eea
where in the last expression, we have assumed that $\Lambda\gg 1$, and set the upper bound of the integral to infinity. As before, the third integral can be analysed by performing the change of integration variable $t = v+1 - \Lambda$. Assuming that $\Lambda\gg 1$, a leading-order expansion in $1/y$ leads to
\bea
\mathcal{F}_3(y,\OmegaPBHf) \propto \frac{\OmegaPBHf^{4/3}}{y^6}\, .
\eea
Recalling that $\OmegaPBHf$ and $\Lambda$ are related through \Eq{Lambda definition}, this implies that $\mathcal{F}_3 \propto y^{-8} \Lambda^{-2}$. The third integral is therefore suppressed by a factor $\Lambda^{-2}$ compared to the first two, and since we have assumed that $\Lambda\gg 1$, the overall integral is dominated by $\mathcal{F}_1$ and $\mathcal{F}_2$, leading to 
\bea
\label{eq:double:integral:y>>1}
{\cal F}(y\gg 1,\OmegaPBHf) \simeq \frac{50625 \pi^2}{2048y^8}\, .
\eea
\bibliographystyle{JHEP}
\bibliography{GW_PBH}
\end{document}